\documentclass
[prb,twocolumn,showpacs,preprintnumbers,amsmath,amssymb]{revtex4}%
\usepackage{graphicx}
\usepackage{latexsym}
\usepackage{amsmath}
\usepackage{graphics}
\usepackage{amssymb}
\usepackage{layout}
\usepackage{verbatim}
\usepackage{amsfonts,epsfig}
\usepackage{slashed}
\usepackage{feynmp}
\usepackage{float}
\usepackage{bm}
\usepackage{bbm}
\usepackage{xcolor}
\usepackage{comment}
\usepackage{amsfonts}%
\providecommand{\U}[1]{\protect\rule{.1in}{.1in}}

\newcommand{\beq}{\begin{equation}}
\newcommand{\eeq}{\end{equation}}
\newcommand{\bea}{\begin{eqnarray}}
\newcommand{\eea}{\end{eqnarray}}

\begin{document}
\title{Robustness of error-suppressing entangling gates in cavity-coupled transmon qubits}
\author{Xiu-Hao Deng, Edwin Barnes, and Sophia E. Economou}
\affiliation{Department of Physics, Virginia Tech, Blacksburg, Virginia 24061, USA}

\begin{abstract}
Superconducting transmon qubits comprise one of the most promising platforms
for quantum information processing due to their long coherence times and to
their scalability into larger qubit networks. However, their weakly anharmonic
spectrum leads to spectral crowding in multiqubit systems, making it
challenging to implement fast, high-fidelity gates while avoiding leakage
errors. To address this challenge, we use a protocol known as SWIPHT [Phys.
Rev. B 91, 161405(R) (2015)], which yields smooth, simple microwave pulses
designed to suppress leakage without sacrificing gate speed through spectral
selectivity. Here, we determine the parameter regimes in which SWIPHT is
effective and demonstrate that in these regimes it systematically produces
two-qubit gate fidelities for cavity-coupled transmons in the range
99.6\%-99.9\% with gate times as fast as 23 ns. Our results are obtained from
full numerical simulations that include current experimental levels of
relaxation and dephasing. These high fidelities persist over a wide range of
system parameters that encompass many current experimental setups and are
insensitive to small parameter variations and pulse imperfections.

\end{abstract}

\pacs{03.67.Lx 03.67.Bg 85.25.Cp }
\maketitle


\section{Introduction}

Rapid progress in the coherence and control of superconducting qubits over the
past decade has made them a frontrunner in the quest for viable quantum
computing platforms.
\cite{Clarke_Nature08,You_Nature11,Devoret_Science13,Martinis_arXiv14} High
fidelity single- and multi-qubit operations,
\cite{Chow_PRL12,Chow_NJP13,Barends_Nature14,DiCarlo_nature10,Fedorov_Nature12}
 as well as initial demonstrations of algorithms and error-correcting codes,
\cite{DiCarlo_nature09,Mariantoni_science11,Lucero_NP12,Chow_NC14,Reed_nature12}
 have been implemented in several multi-qubit devices, and coherence times on
the order of several tens of microseconds and above are now achieved regularly.
\cite{McKay_PRApplied16, Sheldon_PRA16, Corcoles_NC15, Liu_PRX16,
Bultink_arxiv16, Berger_NC15} Perhaps the most promising of these are
transmon qubits, in which insensitivity to charge noise is achieved by
reducing the capacitive energy relative to the Josephson energy through the
use of a large shunt capacitor, leading to a flattening of the charge
dispersion of the energy levels.\cite{Koch_PRA07,Bishop_thesis10,
Barends_PRL13}

There are two general approaches to implementing two-qubit gates in
superconducting qubits. For tunable qubits such as 2D transmons
\cite{Koch_PRA07} or Xmons,\cite{Barends_PRL13} DC magnetic fields are used
to set qubit energies and other circuit parameters. In many systems, such
fields are also used to implement gates by temporarily bringing the system to
a special parameter regime (e.g., a two-qubit resonance), where it is held
idle until different states accumulate the relative phases appropriate for a
desired operation.\cite{yamamoto2010quantum, bialczak2011fast,
dewes2012characterization} The main disadvantage of this
approach is the reliance on flux-tunable qubits, which can have reduced
coherence times due to flux noise.\cite{yoshihara2006decoherence}

The second general approach to gate implementation is to drive one or more
qubits with modulated AC microwave pulses. This method leads to less noise
since the qubit energies are held fixed, and it is the only option for systems
with non-tunable qubits.\cite{paik2016experimental, poletto2012entanglement,
chow2011simple, leek2009using, Majer_Nature07, de2010selective,
rigetti2010fully, de2012selective} The primary challenge with this approach
stems from spectral crowding: a system of several coupled, weakly anharmonic
qubits such as transmons possesses a dense energy spectrum with many closely
spaced transitions. Faster gates are generally preferred since they allow for
faster algorithms. However, faster pulses have broader bandwidth and can thus
lead to the unintended excitation of transitions that are nearly degenerate
with the target transition(s), causing phase and leakage errors. On the other
hand, using spectrally narrower, slower pulses to avoid this problem increases
exposure to relaxation and decoherence. To date, there have been several works
that address this problem in the context of single-qubit gates by devising
pulses that avoid the harmful transitions, either by numerical pulse shaping
\cite{Rebentrost_09} or by engineering the pulse spectrum to contain sharp
holes at the frequencies of the unwanted transitions.
\cite{Motzoi_PRL09,Gambetta_PRA11, Motzoi_PRA13, schutjens2013single,Theis_PRA16} Recent experiments
implementing microwave-driven two-qubit entangling gates in transmon devices
have reported gate times and fidelities ranging from $300-500$ ns and
$87-97\%$.\cite{Chow_NJP13,Chow_NC14,Corcoles_NC15} While there has been recent progress in designing fast leakage-suppressing two-qubit gates using numerical pulse shaping,\cite{Kirchhoff_arxiv17} there remains a need for fast high-fidelity gates based on simple pulses.

Instead of attempting to avoid harmful unwanted transitions, two of us
proposed a new protocol called SWIPHT \cite{Economou_PRB15} to achieve fast,
high-fidelity two-qubit gates by purposely driving the nearest harmful
transition such that the corresponding subspace undergoes trivial cyclic
evolution. This minimizes leakage errors and significantly enhances gate
fidelities without resorting to slow, spectrally selective pulses. While
Ref.~\onlinecite{Economou_PRB15} demonstrated the efficacy of SWIPHT for a set of
typical parameters, a full examination of its regime of validity and its
robustness to parameter variations and decoherence and relaxation has yet to
be carried out.

In this paper, we fill this gap by providing a detailed investigation of the
robustness of the SWIPHT protocol for two-qubit \textsc{cnot} gates. We show
that there exist wide fidelity plateaus in the qubit-frequency landscape where
the fidelity remains above 99.9\%. We also find that with our method, we are
able to maintain the \textsc{cnot} fidelity at 99.9\% while decreasing the
gate time to tens of nano-seconds by exploiting resonances between ground and
excited state transitions. We further demonstrate the robustness of these
results to decoherence and relaxation, variations in qubit-cavity couplings
and qubit frequencies, and pulse deformations using experimentally realistic
decay times and parameter uncertainties.

\section{Analytical approach to gate design}

We consider two transmons coupled to a superconducting cavity. The Hamiltonian
of this system is
\begin{equation}
H_{0}=\omega_{c}a^{\dagger}a+\sum_{\ell=1,2}[\omega_{\ell}b_{\ell}^{\dagger
}b_{\ell}+\frac{\alpha_{\ell}}{2}b_{\ell}^{\dagger}b_{\ell}(b_{\ell}^{\dagger
}b_{\ell}-1)+g_{\ell}(a^{\dagger}b_{\ell}+ab_{\ell}^{\dagger})].
\end{equation}
Here $a^{\dagger}(a)$, $b_{1,2}^{\dagger}(b_{1,2})$ are creation
(annihilation) operators for the cavity and transmons, respectively,
$\omega_{1,2}$ denote the energy splittings between the lowest two states of
each transmon, $\alpha_{1,2}$ are the anharmonicities, and $g_{1,2}$ are the
coupling strengths between each transmon and the cavity. Working in the Fock
basis $\{\left\vert n,i,j\right\rangle \}$, where $n$ is the number of cavity
photons, and $i,j$ denote the energy levels of transmon 1 and 2, respectively,
we diagonalize $H_{0}$ to obtain the dressed eigenstates. In the dispersive
regime and with $g_{1,2}\ll\{\omega_{c},\omega_{1,2}\}$, each dressed state
has a large overlap with one of the bare Fock states; hence, we use $n,i,j$ to
denote the dressed states, but with an additional tilde:
$\{\widetilde{\left\vert n,i,j\right\rangle }\}$.

We define our computational two-qubit states to be the dressed states
$\{\widetilde{\left\vert 000\right\rangle }$, $\widetilde{\left\vert
001\right\rangle }$, $\widetilde{\left\vert 010\right\rangle }$,
$\widetilde{\left\vert 011\right\rangle }\}$, which are very close to the bare
states, $\{\left\vert 000\right\rangle $, $\left\vert 001\right\rangle $,
$\left\vert 010\right\rangle $, $\left\vert 011\right\rangle \}$, for typical
system parameters. The splittings between the bare states $\left|  000\right>
,\left|  001\right>  $ and between $\left|  010\right>  ,\left|  011\right>  $
are equal, as are those between $\left|  000\right>  ,\left|  010\right>  $
and between $\left|  001\right>  ,\left|  011\right>  $. These degeneracies
are slightly broken in the dressed states due to the finite couplings
$g_{1,2}$, allowing one to perform two-qubit entangling gates by driving only
one transition, e.g., driving the $\widetilde{\left|  000\right>
}\Leftrightarrow\widetilde{\left|  010\right>  }$ transition can implement a
\textsc{cnot} gate:
\begin{align}
\hbox{\sc cnot}  &  =e^{i\phi_{a}}\widetilde{\left\vert 000\right\rangle
}\widetilde{\left\langle 010\right\vert }+e^{i\phi_{b}}\widetilde{\left\vert
010\right\rangle }\widetilde{\left\langle 000\right\vert }\nonumber\\
&  +e^{i\phi_{c} }\widetilde{\left\vert 001\right\rangle }%
\widetilde{\left\langle 001\right\vert }+e^{i\phi_{d}}\widetilde{\left\vert
011\right\rangle }\widetilde{\left\langle 011\right\vert }. \label{cnot}%
\end{align}
Here, we generalize the standard \textsc{cnot} by including arbitrary phases
$\phi_{\mu}$; this generalized \textsc{cnot} is maximally entangling and is locally equivalent to the
standard {\sc cnot}. In particular, the two are related by single-qubit $Z$ gates, which have recently been experimentally demonstrated for fixed-frequency qubits.\cite{Ku_arxiv17,McKay_arxiv16}

The \textsc{cnot} gate in Eq.~\eqref{cnot} can be implemented by driving only
the first transmon with a microwave $\pi$-pulse that is resonant with the
$\widetilde{\left\vert 000\right\rangle }\Leftrightarrow\widetilde{\left\vert
010\right\rangle }$ transition. The total Hamiltonian can be written in the
bare eigenbasis as
\begin{equation}
\mathcal{H}(t)=H_{0}+b_{1}\Omega(t)e^{i\omega_{p}t}+b_{1}^{\dagger}%
\Omega(t)e^{-i\omega_{p}t},
\end{equation}
where $\Omega(t)$ and $\omega_{p}$ are the pulse envelope and frequency,
respectively. In the dispersive regime, the simplest way to ensure that this
transition is the only one excited by the pulse is to use a very narrowband
pulse---an approach which necessarily leads to long gate times. To avoid
making this sacrifice in gate speed, we instead employ the SWIPHT method
\cite{Economou_PRB15,Barnes_arxiv16} to remove the effects of inadvertently
driving unwanted transitions without resorting to spectrally narrow, slow
pulses. For typical experimental values of the qubit-cavity couplings
$g_{1,2}$, there is exactly one nearest \textquotedblleft
harmful\textquotedblright\ transition, namely the $\widetilde{\left\vert
001\right\rangle }\Leftrightarrow\widetilde{\left\vert 011\right\rangle }$
transition, which competes with the target transition, $\widetilde{\left\vert
000\right\rangle }\Leftrightarrow\widetilde{\left\vert 010\right\rangle }$.
The SWIPHT protocol calls for purposely driving this transition in such a way
that the net evolution operator in this subspace is proportional to an
identity operation.

\begin{figure}[ptb]
\includegraphics[width=0.6\columnwidth]{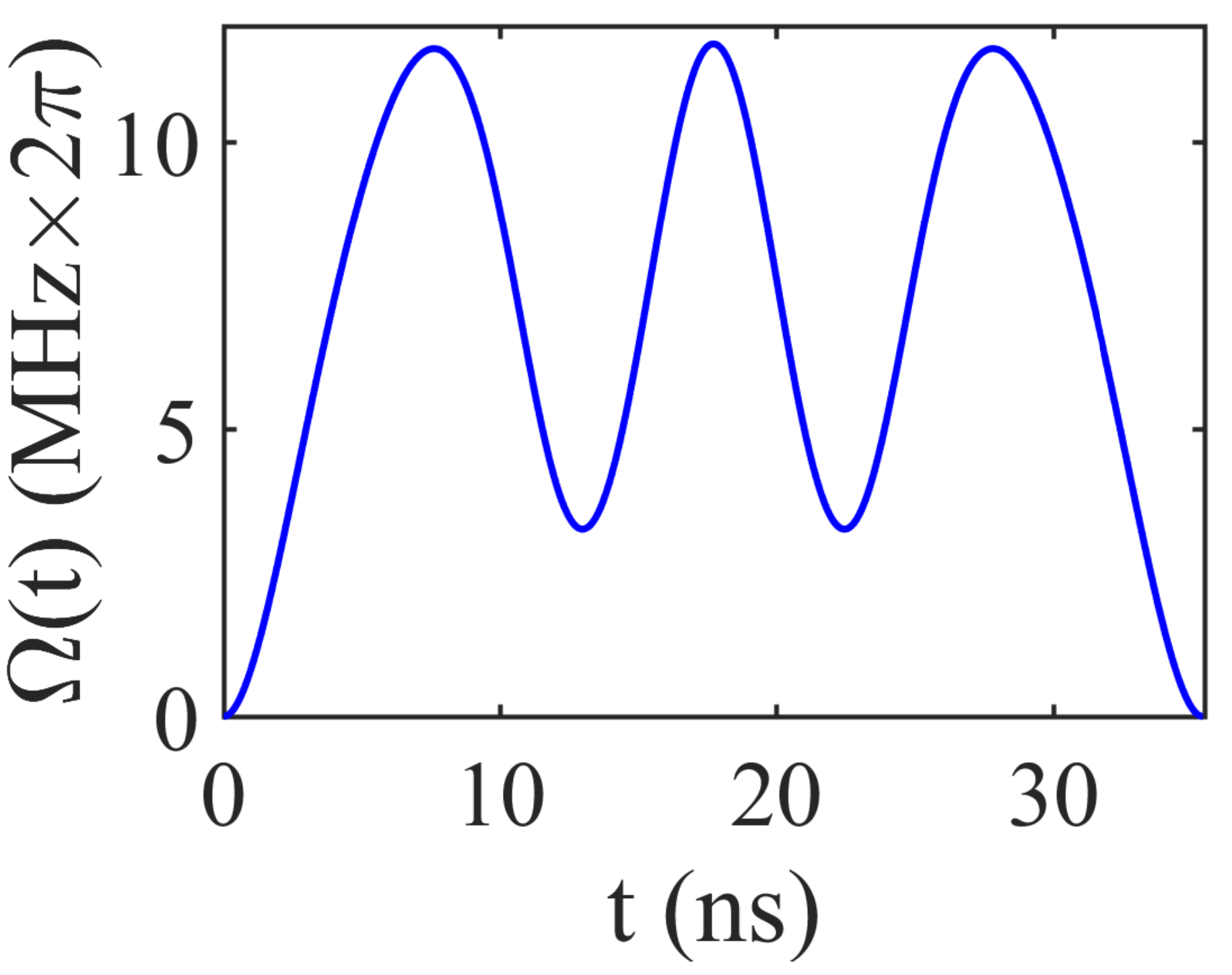}\caption{Pulse envelope from
Ref.~\onlinecite{Economou_PRB15} that implements a \textsc{cnot} gate in 35.4 ns
with $\omega_{1}=6.2$ GHz, $\omega_{2}=6.8$ GHz, $\omega_{c}=7.15$ GHz,
$\alpha_{1,2}=350$ MHz, $g_{1,2}=250$ MHz.}%
\label{fig:pulse}%
\end{figure}

\begin{figure*}[t]
\includegraphics[width=1.6\columnwidth]{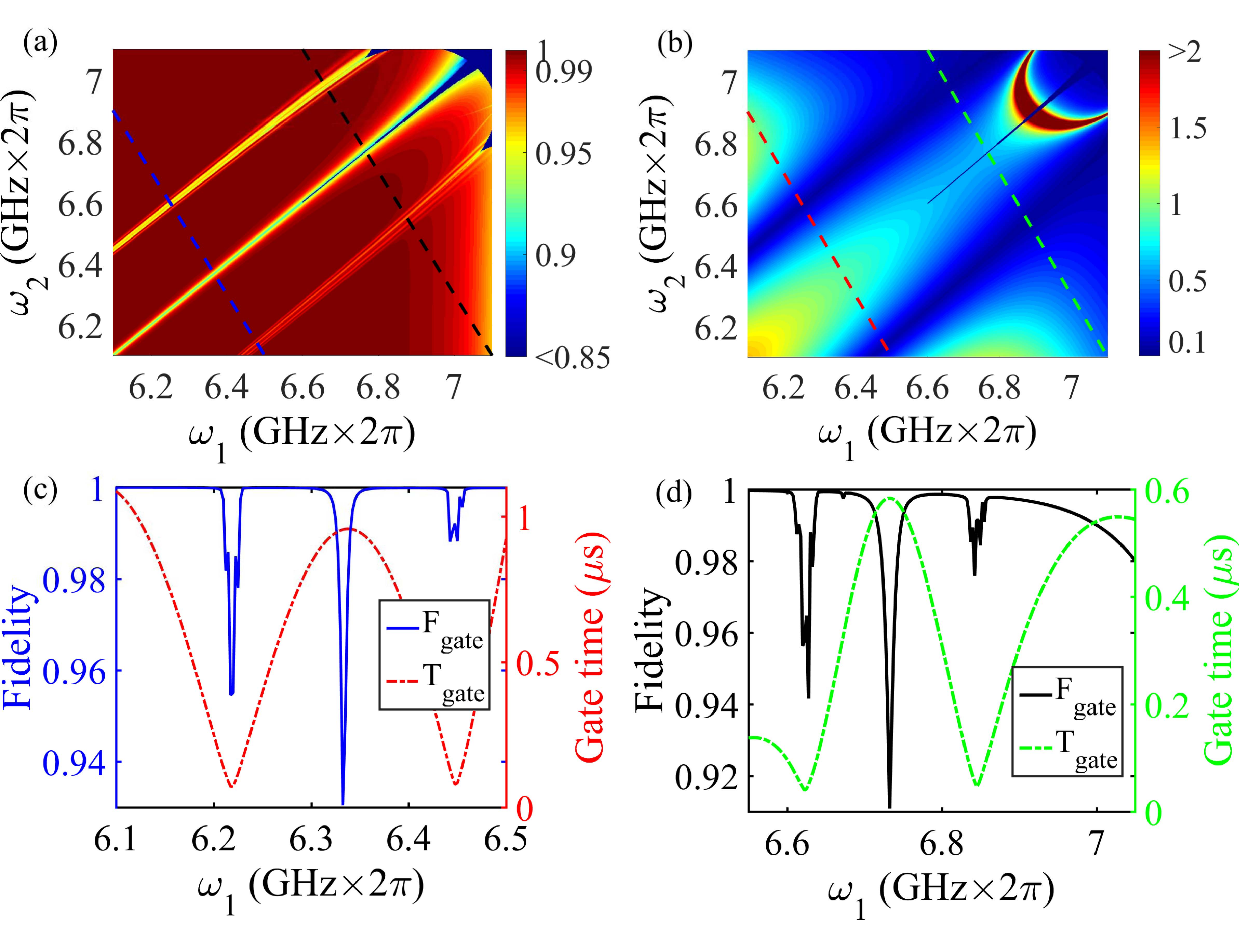}\caption{(a) \textsc{cnot}
fidelity and (b) gate time ($\mu s$) versus qubit
frequencies. $g_{1,2}=100$ MHz, while all other parameters are as in
Fig.~\ref{fig:pulse}. Cross-section plots for the four dashed lines are shown
in (c) and (d) with corresponding colors.}%
\label{Fig_FidelityQEn}%
\end{figure*}

In the computational two-qubit subspace spanned by the states
$\widetilde{\left\vert 000\right\rangle }$, $\widetilde{\left\vert
010\right\rangle }$, $\widetilde{\left\vert 001\right\rangle }$,
$\widetilde{\left\vert 011\right\rangle }$ (note the unconventional basis
ordering), the Hamiltonian of the driven transmon-cavity system is
approximately
\[
H_{cs}{\approx}\left(
\begin{array}
[c]{cccc}%
-E/2{-}\epsilon & \Omega(t)e^{i\omega_{p}t} & 0 & 0\\
\Omega(t)e^{-i\omega_{p}t} & E/2{-}\epsilon & 0 & 0\\
0 & 0 & -(E{-}\delta)/2 & \Omega(t)e^{i\omega_{p}t}\\
0 & 0 & \Omega(t)e^{-i\omega_{p}t} & (E{-}\delta)/2
\end{array}
\right)  ,
\]
where $E$ is the energy splitting between $\widetilde{\left\vert
000\right\rangle }$ and $\widetilde{\left\vert 010\right\rangle }$, and
$E-\delta$ is the splitting between $\widetilde{\left\vert 001\right\rangle }$
and $\widetilde{\left\vert 011\right\rangle }$. We have shifted the overall
energy by $-E/2-\epsilon$, where $\epsilon+\delta/2$ is the energy of state
$\widetilde{\left\vert 001\right\rangle }$. We denote the pulse duration by
$\tau_{p}$. We have also neglected the subleading terms in the off-diagonal
$2\times2$ blocks (but not in the simulations).\textbf{ }To implement a SWIPHT
\textsc{cnot} gate, we set $\omega_{p}=E$ and engineer $\Omega(t)$ such that
the evolution operator generated by $H_{cs}$ coincides with the \textsc{cnot}
gate given in Eq.~\eqref{cnot} with $\phi_{\mu}=0$. Matching the form of the
upper left $2\times2$ subspace requires the area of the pulse to be given by
$\int_{0}^{\tau_{p}}dt\Omega(t)=\pi/2$, as is consistent with a $\pi$-pulse.

Engineering the evolution operator in the lower right 2$\times$2 subspace to
be an identity operation at time $t=\tau_{p}$ is more challenging since it is
not possible to analytically solve the Schr$\ddot{\hbox{o}}$dinger equation
for an off-resonant pulse with arbitrary envelope $\Omega(t)$. We can overcome
this difficulty by making use of a partial-reverse engineering formalism
introduced in Refs.~\onlinecite{Barnes_PRL12,Barnes_PRA13}. In
Ref.~\onlinecite{Economou_PRB15}, this formalism was used to obtain the pulse shown
in Fig.~\ref{fig:pulse}, which implements a \textsc{cnot} gate with fidelity
$>99$\% in 35.4 ns. A brief review of the construction of this pulse is given in Appendix~\ref{app:SWIPHTreview}. The duration of the pulse is given by $\tau_{p}%
{=}5.87/|\Delta|$, where $\Delta{=}\omega_{p}{-}(E{-}\delta)$ is the detuning
of the pulse relative to the harmful transition. For $\omega_{p}{=}E$ we have
$\Delta{=}\delta$, and thus $\tau_{p}$ depends on the system parameters
through the dependence on the transition frequency difference $\delta$, which
is due to the cavity-mediated coupling. For the parameters considered in
Ref.~\onlinecite{Economou_PRB15} (summarized in the caption of Fig.~\ref{fig:pulse}%
), $\delta=26.4$ MHz.

\section{Numerical results and robustness}

\subsection{Dependence of gate fidelity on qubit frequencies}

\begin{figure*}[t]
\centering
\includegraphics[width=0.75\columnwidth]{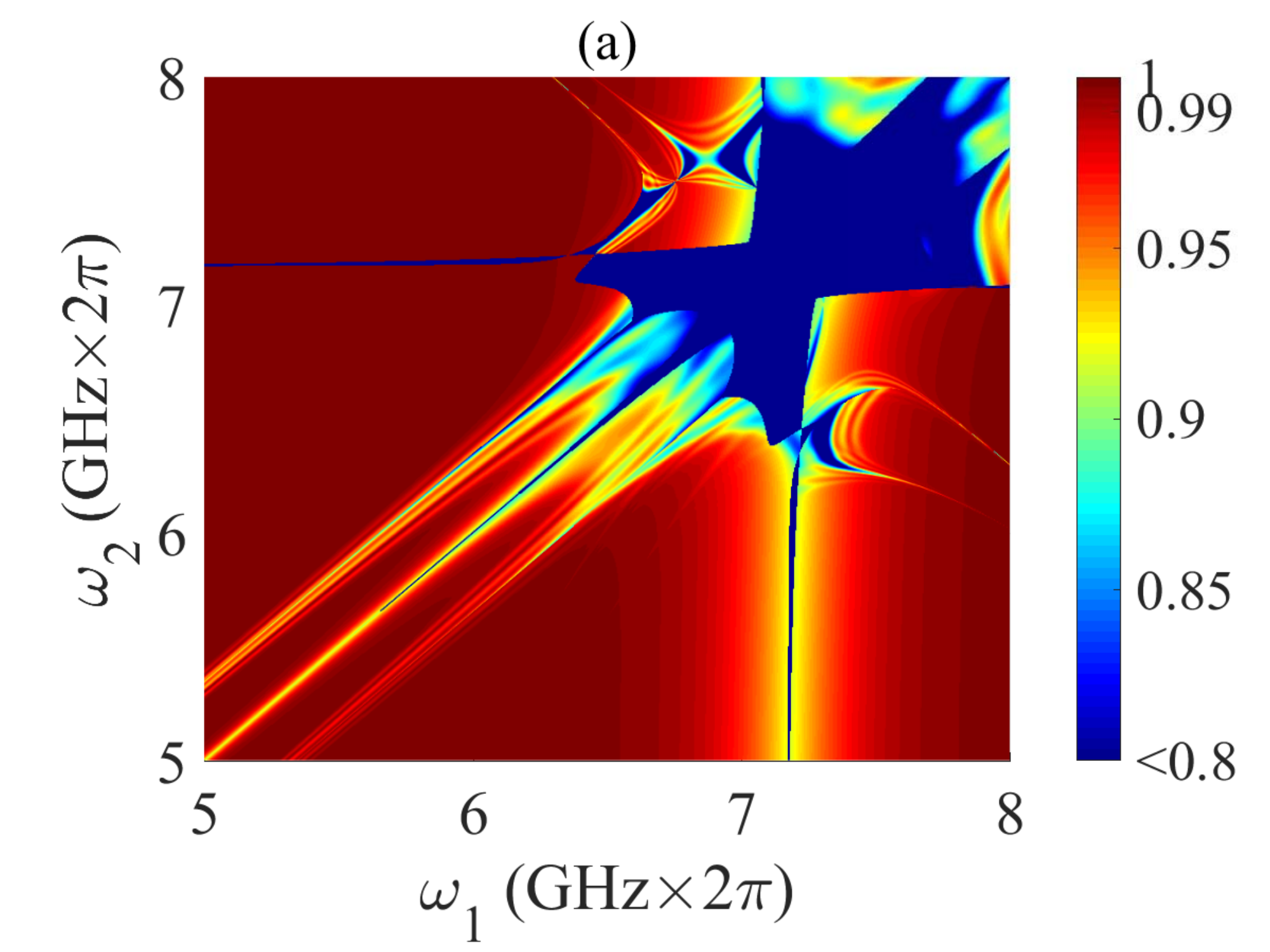}
\includegraphics[width=0.75\columnwidth]{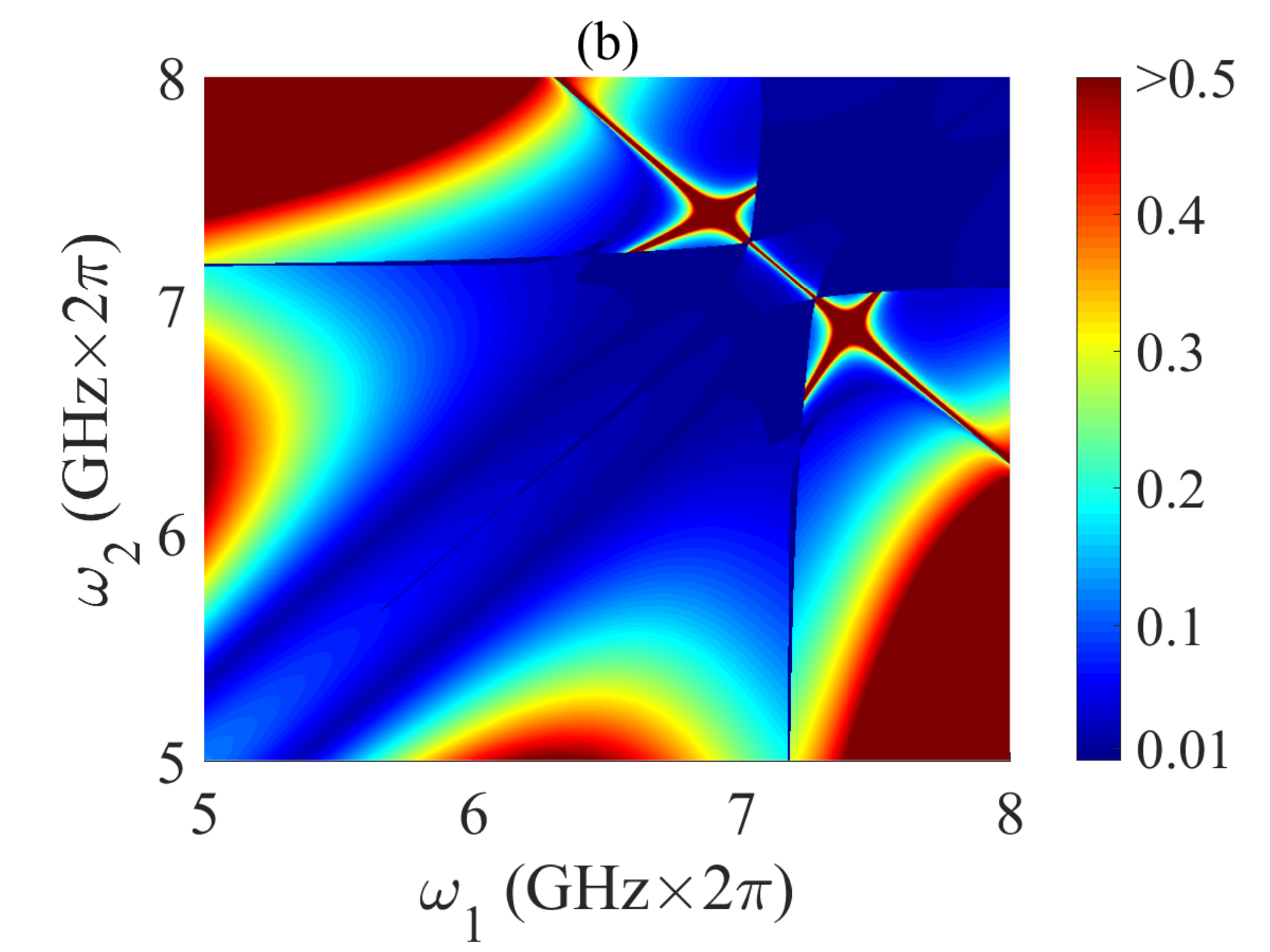}
\caption{\textsc{cnot} fidelity (a) and gate time ($\mu s$) (b) versus qubit frequencies for $g_{1,2}=250$ MHz.}%
\label{Fig_Fidelitye1e2}%
\end{figure*}

First, we study the dependence of the \textsc{cnot} fidelity and gate speed on
the transmon frequencies. For the moment, we neglect relaxation and dephasing,
although these effects will be included below. In this case, we define the
gate fidelity as in Ref.~\onlinecite{Pedersen_PLA07}, which accounts for leakage
outside the computational two-qubit subspace:
\begin{equation}
\mathcal{F}^{gate}=\frac{1}{20}\left[  \hbox{Tr}(MM^{\dagger}%
)+|\hbox{Tr}(M)|^{2}\right],
\end{equation}
where $M=U_{ideal}U^{\dagger}$, and $U$ is the actual evolution operator,
while $U_{ideal}$ is the target gate operation, here taken as the
\textsc{cnot} gate defined in Eq.~\eqref{cnot}. We solve the time-dependent
Schr\"{o}dinger equation for the evolution operator generated by our
analytical SWIPHT pulse keeping three cavity and four transmon states, for a
total of 48 states. The number of states was increased until convergence in
the results was achieved. For each set of system parameters, we optimize over
the phases $\phi_{\mu}$. Our numerical results for $\mathcal F^{gate}$ and $\tau_{p}$
are shown in Fig.~\ref{Fig_FidelityQEn}. The most important features of
Fig.~\ref{Fig_FidelityQEn}(a) are the large plateaus where $\mathcal{F}%
^{gate}$ is well above 0.999 (dark red); these occur in regions where
$\omega_{1,2}$ are detuned from the three sharp linear features evident in the
figure. The central feature corresponds to the qubit-qubit resonance,
$\omega_{1}=\omega_{2}$, while the two \textquotedblleft
secondary\textquotedblright\ resonances occur where $\omega_{1}=\omega
_{2}+\alpha_{1}$ or $\omega_{1}=\omega_{2}-\alpha_{2}$, corresponding to the
$\left\vert 0\right\rangle \Leftrightarrow\left\vert 1\right\rangle $
transition of one qubit becoming degenerate with the $\left\vert
1\right\rangle \Leftrightarrow\left\vert 2\right\rangle $ transition of the
other. Near these resonances, additional harmful transitions become important,
causing a decrease in fidelity. Further details regarding these resonances can be found in Appendix~\ref{app:secondaryresonances}. This figure also exhibits an asymmetry between
the dependencies on $\omega_{1,2}$ caused by the fact that only transmon 1 is
driven. Since the high-fidelity plateau is broader for $\omega_{1}<\omega_{2}%
$, we see that it is more advantageous to drive the transmon that is further
detuned from the cavity.

Fig.~\ref{Fig_FidelityQEn}(b) reveals that there is significant overlap
between the high-fidelity plateaus and the parameter regions where the gate
times are below 150 ns (blue). The fastest pulses occur near the secondary
resonances because these give rise to a larger splitting, $\delta$, between
the target and harmful transitions, which in turn reduces the SWIPHT gate time
since $\tau_{p}\sim1/|\delta|$. Further details can be found in Appendix~\ref{app:secondaryresonances}. Figs.~\ref{Fig_FidelityQEn}(c),(d) show the \textsc{cnot} fidelity and gate time along two one-dimensional slices in qubit-frequency space. Importantly,
we see that while the fidelity quickly increases up to above 0.999 as
$\omega_{1}$ is tuned away from a secondary resonance, the gate time increases
more slowly. Thus, the best combination of low gate time and high fidelity is
achieved when the system lies close to a secondary resonance. From
Fig.~\ref{Fig_FidelityQEn}(d), which shows a slice closer to the cavity
frequency, $\omega_{c}=7.15$ GHz, we in fact see that as $\omega_{1}$ is
reduced, the gate time saturates at around 150 ns while the fidelity continues
to improve. Below, we show that the gate time can be further reduced by more than a factor of 6 by adjusting system and pulse parameters appropriately.

Fig.~\ref{Fig_Fidelitye1e2} shows zoomed-out versions of Figs. 2(a),(b), where the full extent of the broad high-fidelity plateaus is more evident.

\subsection{Performance under relaxation and dephasing}

\begin{figure}[ptb]
\includegraphics[width=0.67\columnwidth]{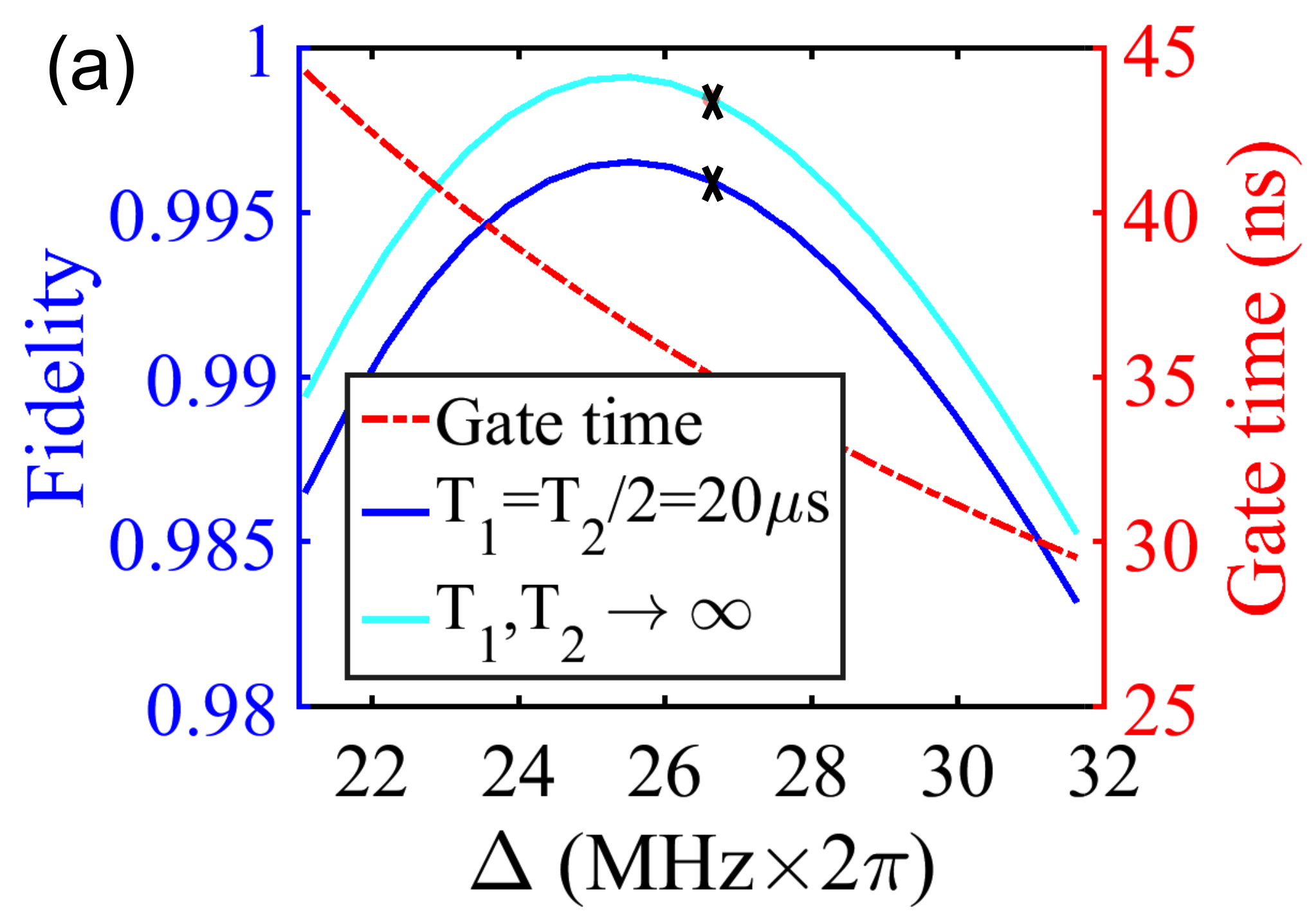}
\includegraphics[width=0.67\columnwidth]{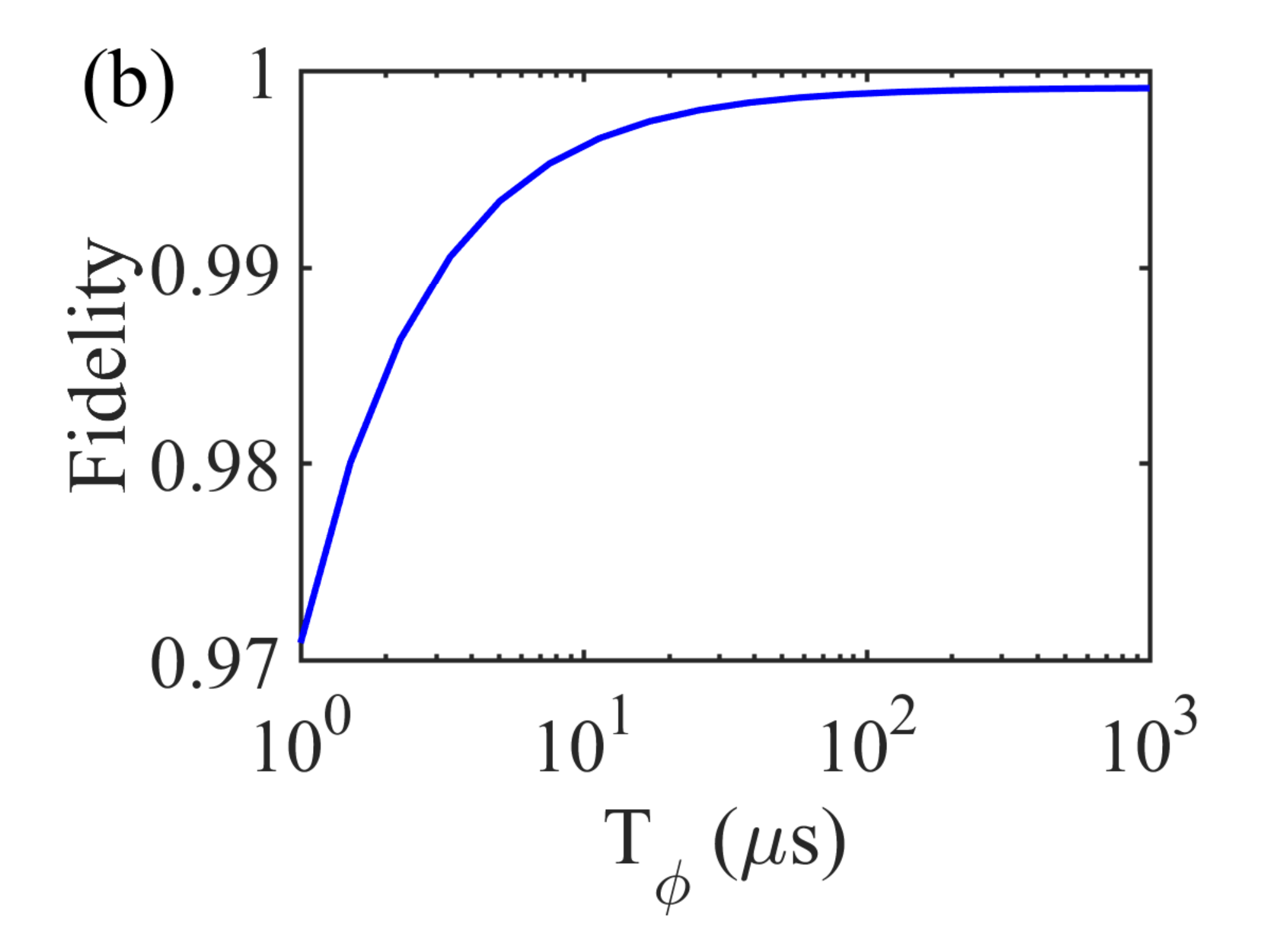}
\caption{(a) Fidelity versus
detuning $\Delta$ for system parameters in Fig.~\ref{fig:pulse}. Cyan
line is the fidelity for noiseless case while blue line is for $T_{1}%
=T_{2}=T_{\phi}/2=20\mu s$. The black crosses indicate $\Delta=\delta$. Dashed
red line shows gate time. (b) Average
fidelity versus dephasing time $T_{\phi}$ for $\Delta=25.5$ MHz and pulse duration is 36.6 ns, which are the optimal values found in (a).}
\label{fig:dephasing}%
\end{figure}

Next, we evaluate the impact of relaxation and decoherence on our gate by solving the Lindblad equation:
\begin{equation}
\dot{\rho}=i[\rho,\mathcal{H}(t)]+\sum_{\ell=1,2}\left(  \frac{1}{T_{1}%
}\mathcal{D}[b_{\ell}]+\frac{1}{T_{\phi}}\mathcal{D}[b_{\ell}^{\dagger}%
b_{\ell}]\right)  , \label{lindblad}%
\end{equation}
with $\mathcal{D}[L]=L\rho L^{\dagger}-\frac{1}{2}\{L^{\dagger}L,\rho\}$. The
first Lindblad term corresponds to qubit relaxation (time scale $T_{1}$),
while the second corresponds to pure dephasing (time scale $T_{\phi}$) caused
by charge fluctuations.\cite{Koch_PRA07,Bishop_thesis10} Here $1/T_{2}%
=1/2T_{1}+1/T_{\phi}$. We have neglected cavity decay in our simulation
because its time scale is typically much larger than $T_{1}$ and $T_{2}$ and
because our gate scheme causes minimal cavity excitation. With noise terms
included, $\mathcal{F}^{gate}$ is no longer a suitable definition of fidelity,
and we instead perform quantum state tomography. We prepare 16 input states in
total, chosen from the set $\{\left\vert 0\right\rangle ,\left\vert
1\right\rangle ,(\left\vert 0\right\rangle +\left\vert 1\right\rangle
)/\sqrt{2},(\left\vert 0\right\rangle +i\left\vert 1\right\rangle )\sqrt{2}\}$
for each qubit. We calculate the average fidelity, defined as $\mathcal{F}%
=\frac{1}{16}\sum_{j=1}^{16}\hbox{Tr}(\rho_{ideal}^{j}\rho_{sim}^{j})$, where
$\rho_{ideal}^{j}$ is the ideal target state, while $\rho_{sim}^{j}$ is the
final density matrix obtained by solving Eq.~\eqref{lindblad}. We again use a
48-state Hilbert space to achieve convergence and optimize over $\phi_{\mu}$.

We first study the dependence of $\mathcal{F}$ on the pulse detuning $\Delta$.
Fig.~\ref{fig:dephasing}(a) shows this dependence with and without noise,
where it is clear that a slight detuning away from the idealized value based
on $H_{cs}$ ($\Delta{=}\delta{=}26.4$~MHz) down to $\Delta{=}25.5$~MHz brings
the fidelity up to 0.9991 without noise or up to 0.9963 with noise for typical
experimental values of $T_{1}$, $T_{2}$. The figure also shows that this
improvement comes with a slight increase in the gate time from 35.4 ns up to
36.6 ns. In Fig.~\ref{fig:dephasing}(b), we show the dependence of the fidelity on $T_{\phi}=2T_1=2T_2$ for the optimized pulse with $\tau_{p}=36.6$ ns, where we find that
$\mathcal{F}\geq0.995$ for $T_{2}=T_{1}\geq7\mu$s. We also examine the
performance of our gate for several sets of measured parameter values taken
from recent experimental works in Appendix~\ref{app:experimentalparameters}. In Appendix~\ref{app:localphases}, we further show that the optimized local phases $\phi_\mu$ that enter into the generalized {\sc cnot} gate (Eq.~\eqref{cnot}) are not sensitive to experimental uncertainties in parameter values.

\subsection{Asymmetry of coupling strength and anharmonicity}

\begin{figure}[ptb]
\includegraphics[width=0.7\columnwidth]{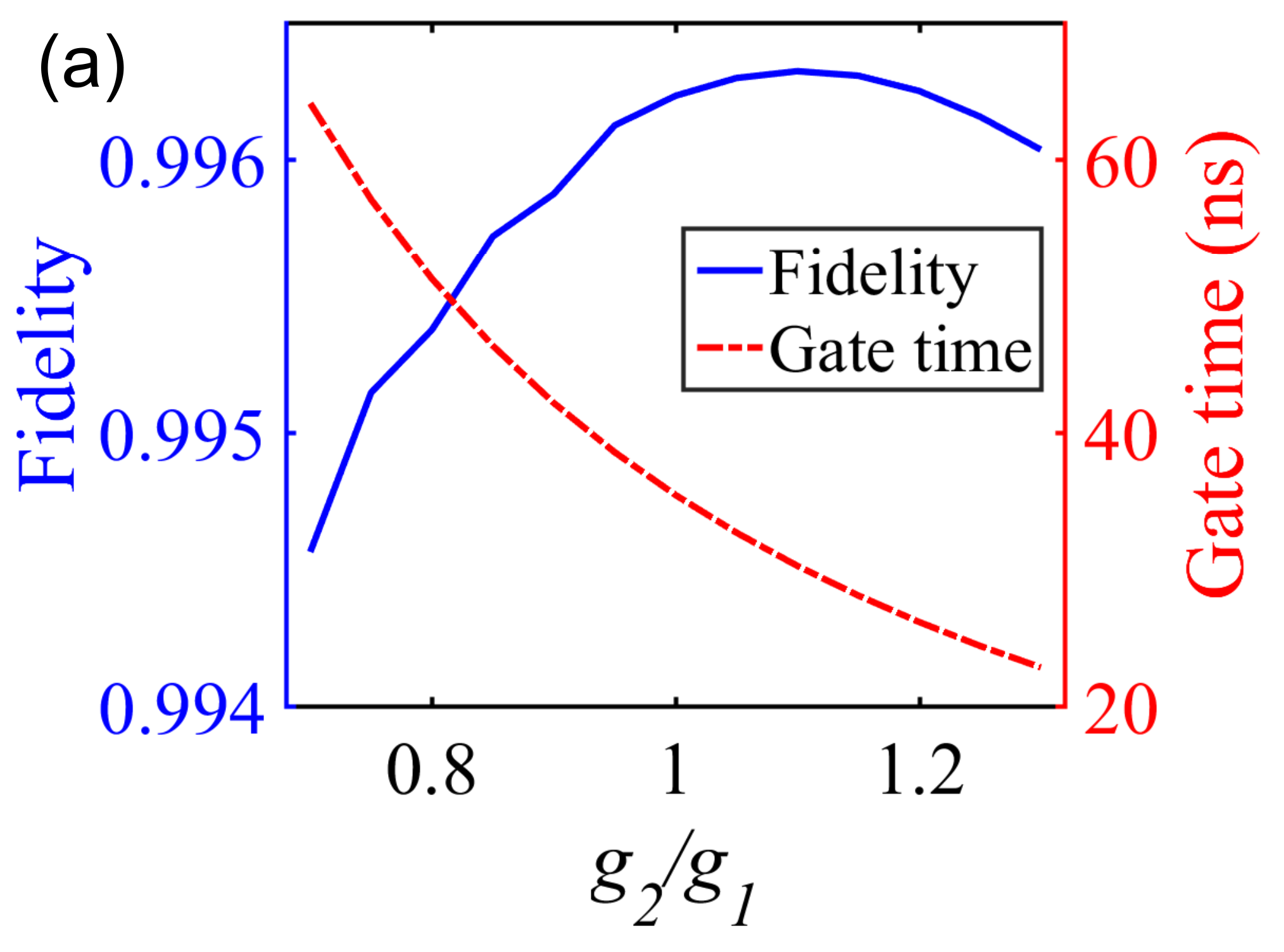}
\includegraphics[width=0.7\columnwidth]{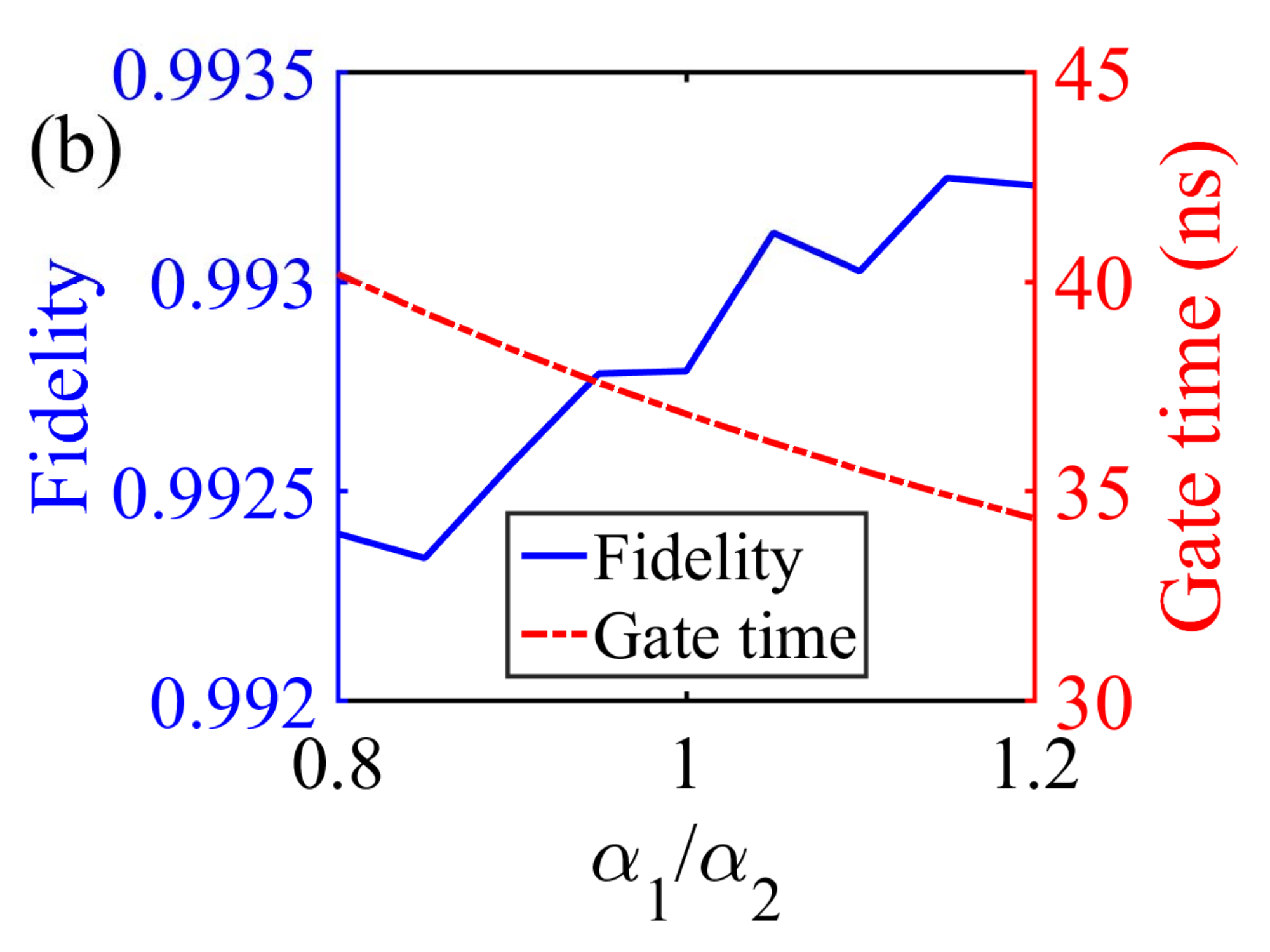}\caption{(a) Fidelity and gate time versus coupling asymmetry and (b) fidelity versus anharmonicity asymmetry. The system parameters are as in Fig.~\ref{fig:pulse} except as shown ($g_2$ and $\alpha_1$ are varied), and $T_{1}=T_{2}=20\mu s$.}%
\label{fig:asymmetry}%
\end{figure}

In a real setup, the qubit-cavity coupling strengths $g_{1}$, $g_{2}$ may
differ. Fig.~\ref{fig:asymmetry}(a) shows that the fidelity remains $>0.995$
even when the couplings differ by more than 20\%. We also find that further
optimization of the gate is possible if the coupling of the undriven transmon
(here $g_{2}$) is tuned to be slightly larger than that of the driven qubit ($g_{1}$). The
figure further shows that the gate time is simultaneously reduced to as low as
23 ns while the fidelity remains above 0.996 even in the presence of relaxation and dephasing ($T_1=T_2=20 \mu$s).

\begin{figure}[ptb]
\includegraphics[width=0.7\columnwidth]{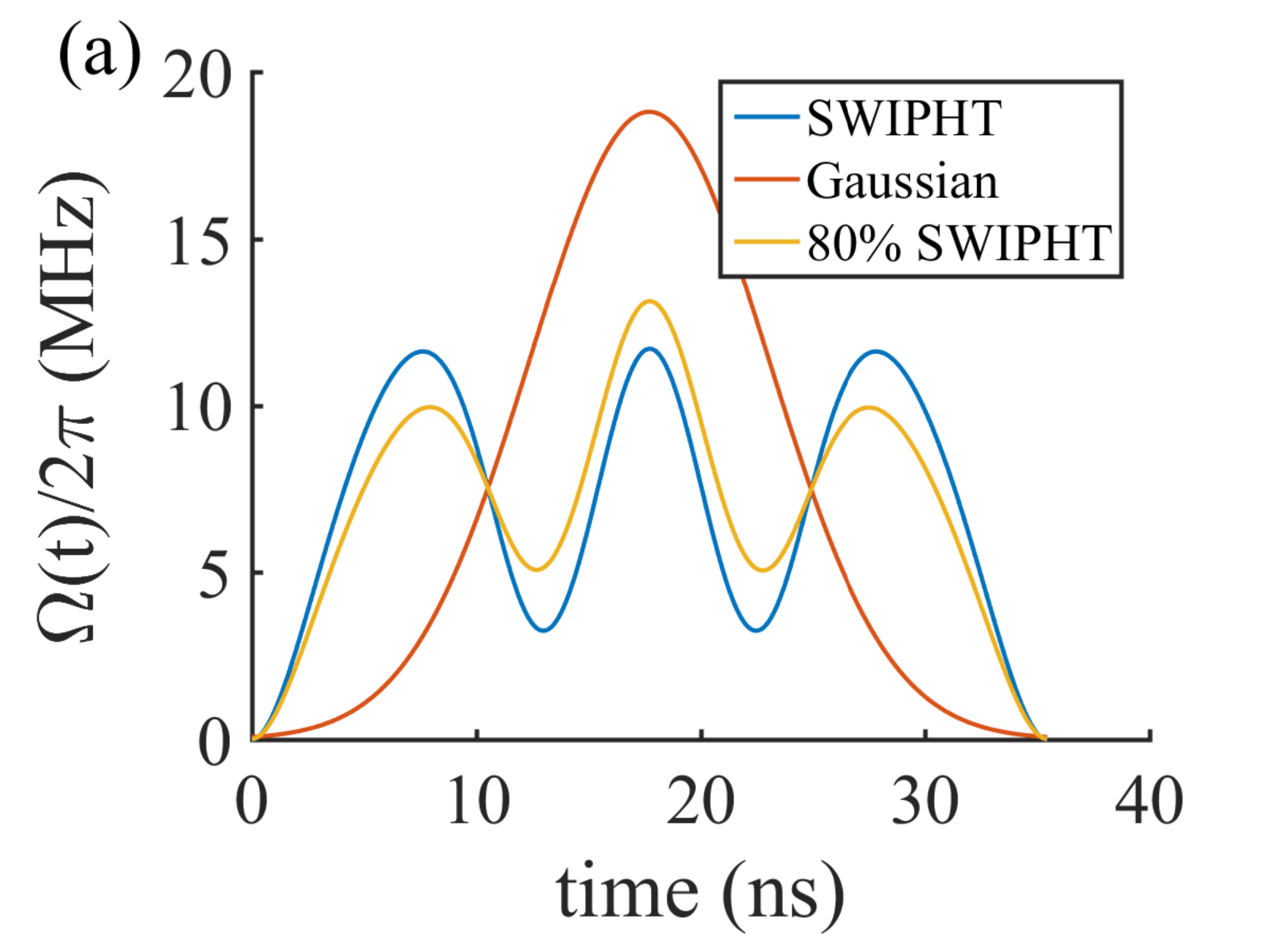}
\includegraphics[width=0.7\columnwidth]{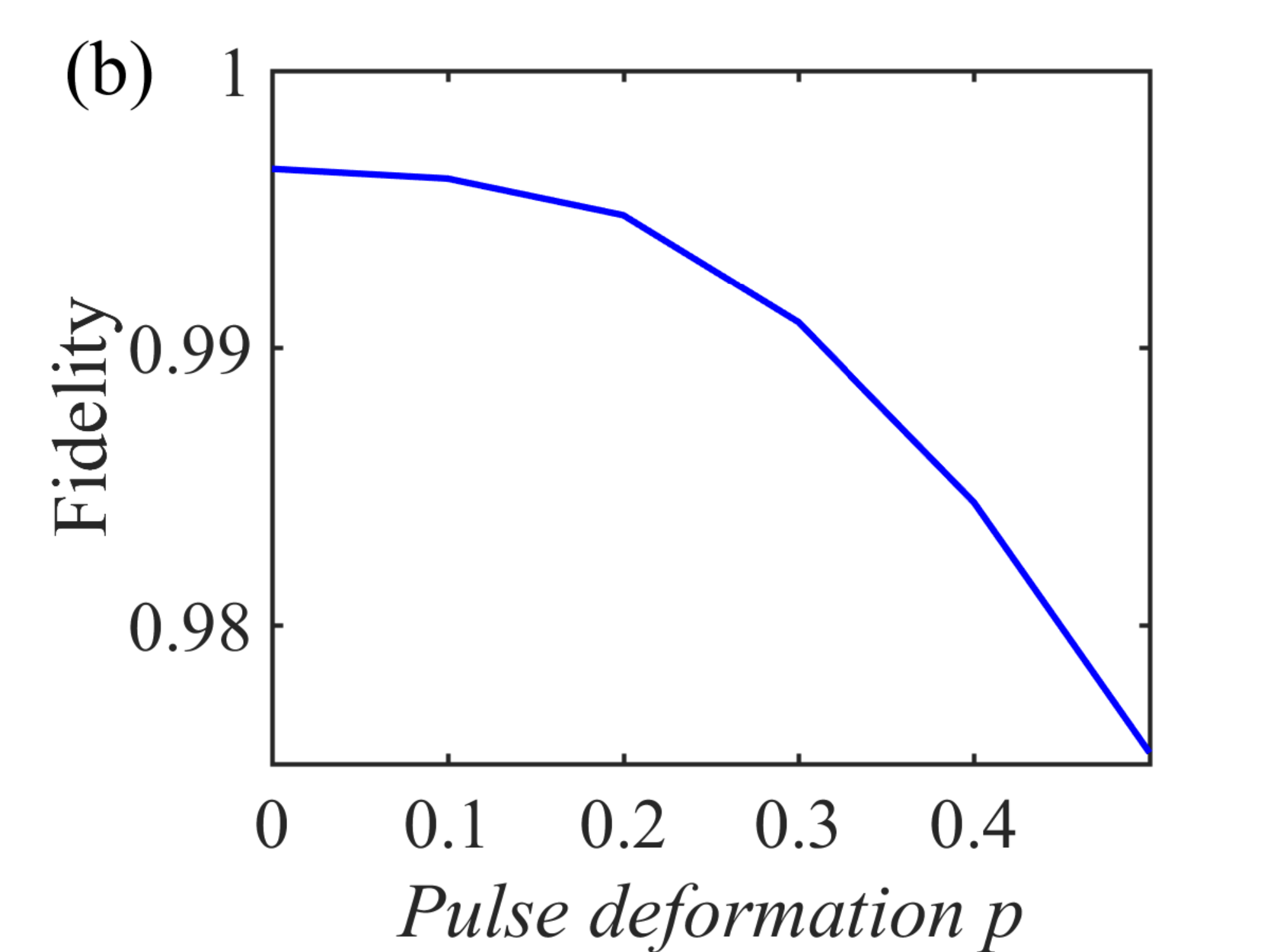}
\includegraphics[width=0.67\columnwidth]{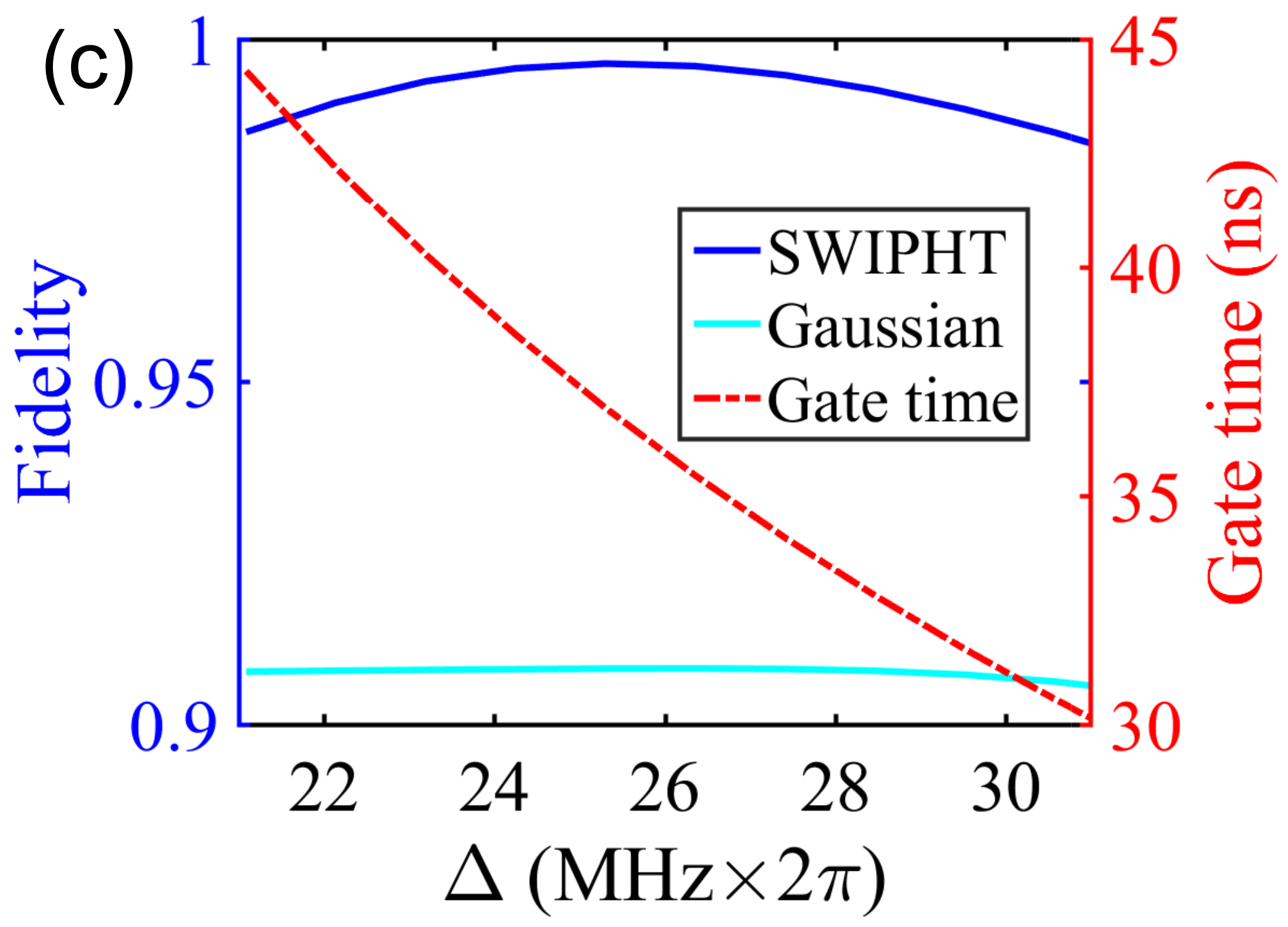}
\caption{(a) Pulses with Gaussian-type pulse deformations as described in
Eq.~\eqref{Eq_PulseDeform} for three different values of the deformation
parameter: $p=0$ (SWIPHT), $p=1$ (Gaussian), and $p=0.2$ (80\% SWIPHT, 20\%
Gaussian). (b) Fidelity versus degree of pulse deformation from Eq.~\eqref{Eq_PulseDeform}. (c) \textsc{cnot} fidelity and gate time versus detuning of the
pulse relative to the harmful transition for the SWIPHT and Gaussian pulses
shown in the left panel. All system and pulse parameters are as in Fig.~\ref{fig:pulse} and $T_1=T_2=20\mu$s.}
\label{fig:deformation}
\end{figure}

SWIPHT is similarly robust against anharmonicity differences. So far, we assumed that
both qubits share the same value of anharmonicity, $\alpha_{1}=\alpha_{2}$,
for simplicity. We have rerun the simulations shown in Fig.~\ref{Fig_Fidelitye1e2} for $\alpha_1=350$ MHz, $\alpha_2=300$ MHz. The results are essentially unchanged from those shown in Fig.~\ref{Fig_Fidelitye1e2} except for a shift in the location of one secondary resonance as follows trivially from the change in $\alpha_2$. In Fig \ref{fig:asymmetry}(b), we show the SWIPHT
\textsc{cnot} gate fidelity versus asymmetry in anharmonicity between the two
transmons. It is clear from the figure that not only is the SWIPHT gate robust
against anharmonicity differences, but that such differences can even lead to
further improvement in the fidelity.

\subsection{Pulse deformation}

Next, we consider the robustness of the results to
Gaussian-type pulse deformations of the form
\begin{equation}
\Omega(t)=(1-p)\ast\Omega_{\text{SWIPHT}}(t)+p\ast\Omega_{\text{Gaussian}}(t),
\label{Eq_PulseDeform}%
\end{equation}
where $\Omega_{\operatorname{SWIPHT}}(t)$ is the pulse shown in
Fig.~\ref{fig:pulse}. The Gaussian pulse, $\Omega_{\text{Gaussian}}$, is chosen to have the same
area ($\pi/2$) and duration (35.4 ns) as the SWIPHT pulse. Explicitly, we use
\begin{equation}
\Omega_{\text{Gaussian}}(t)=A_{G}e^{-(t-\tau_{G}/2)^{2}/(2\tau_{G}^{2})},
\end{equation}
where $A_{G}=2\pi\ast18.8$ MHz, and $\tau_{G}=0.15\ast35.4$ ns. The resulting
pulses for three different values of $p$ are shown in
Fig.~\ref{fig:deformation}(a). Fig.~\ref{fig:deformation}(b) shows the fidelity as a function of $p$ (with relaxation and dephasing included), where it is evident that the gate performance is essentially
unchanged for deformations up to the 10\% level, further highlighting the
robustness of our gate. In Fig.~\ref{fig:deformation}(c), we show a comparison of the SWIPHT and pure
Gaussian pulses; we see that the SWIPHT pulse performs dramatically
better for gate times on the order of a few tens of nanoseconds.

\begin{figure}[ptb]
\centering\includegraphics[width=0.7\columnwidth]{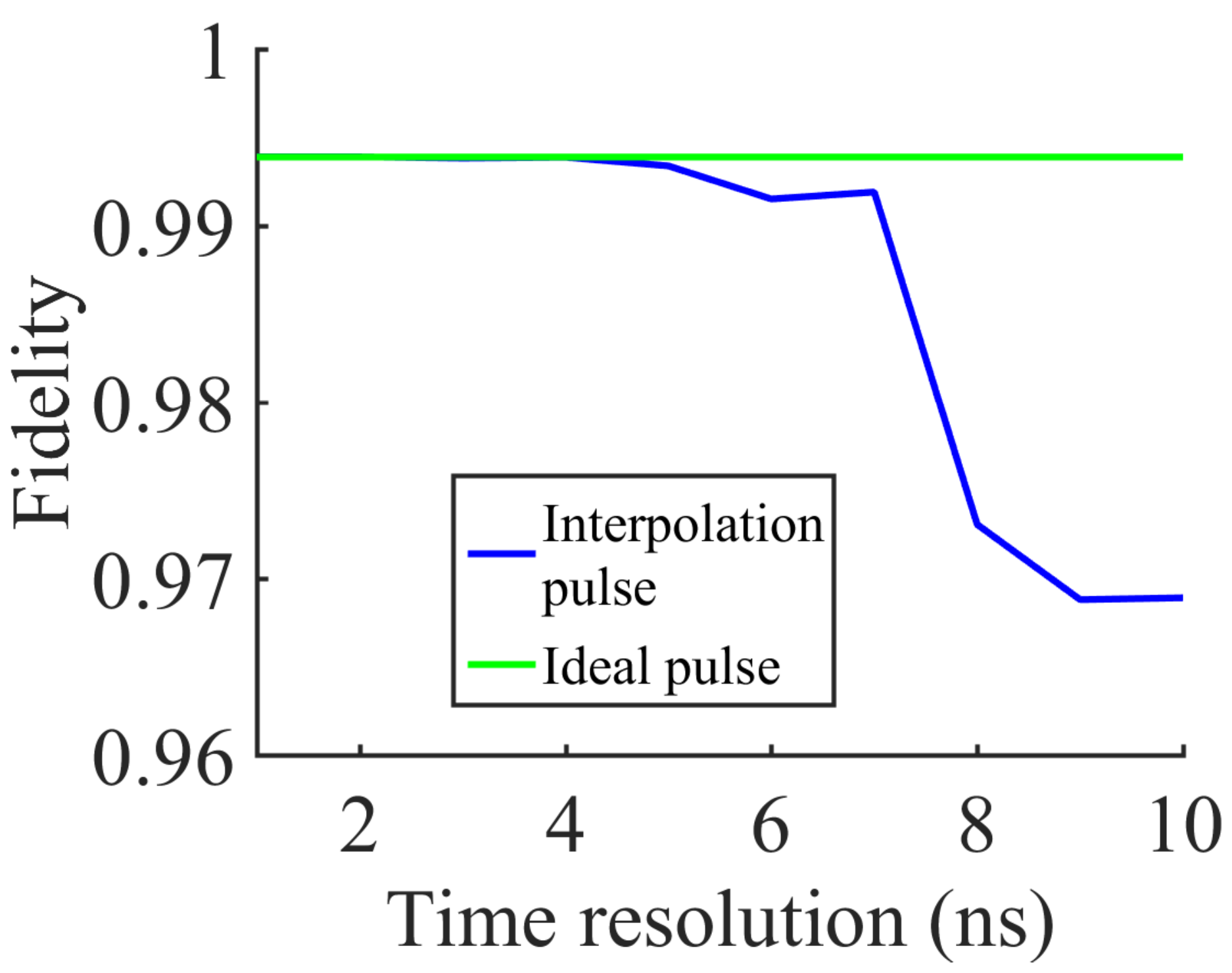}\caption{Fidelity
versus time resolution of pulse envelope. Pulse and system parameters are
as in Fig.~\ref{fig:pulse}, $T_1=T_2=20\mu$s.}
\label{Fig_interpolate_pulses}
\end{figure}

Pulse deformations can also result from the finite time resolution of a
pulse generator. In Fig.~\ref{Fig_interpolate_pulses}, we show the
SWIPHT fidelity versus time resolution. The plateau of fidelity that persists
up to $4$ ns shows that SWIPHT is very robust to these pulse deformations. These findings demonstrate that SWIPHT is effective even
with modest pulse-shaping capabilities.

\section{Conclusion}

In conclusion, we have shown that the SWIPHT method can produce \textsc{cnot}
gates in cavity-coupled transmon systems with fidelities well above 99.5\% and
gate times below 30 ns even when realistic levels of decoherence, relaxation,
and parameter uncertainties are taken into account. In general, we find that
SWIPHT performs well when the degeneracy between target and harmful
transitions is strongly broken, either through strong qubit-cavity couplings,
reduced qubit-cavity detunings, or transition resonances. Our work is of
immediate use to ongoing experimental efforts to optimize the performance of
transmon systems operated with microwave control.

\acknowledgments
We would like to thank A. Lupascu for interesting discussions and helpful comments.

\appendix

\section{SWIPHT pulse shape}\label{app:SWIPHTreview}

In this appendix, we review how the analytical pulse used to implement the SWIPHT
\textsc{cnot} gate is derived.\cite{Economou_PRB15} As described in the main
text, in order to implement this gate, we must design a pulse that implements
a $\pi$ rotation about $x$ on the target transition and an identity operation
on the harmful transition. The former is achieved by making the pulse resonant
with the target transition and choosing the pulse area to be $\int_{0}%
^{\tau_{p}}dt\Omega(t)=\pi/2$. Ensuring that the harmful transition undergoes
a trivial identity operation is more challenging, and we solve this problem by
making use of the formalism introduced in Ref.~[\onlinecite{Barnes_PRA13}] for
analytically solving the time-dependent Schr$\ddot{\hbox{o}}$dinger equation.
In this formalism, analytical solutions are obtained by expressing both the
evolution operator and the driving field in terms of an auxiliary function
$\chi(t)$. One imposes constraints on $\chi(t)$ that ensure the desired
evolution is obtained and then reads off the corresponding driving field that
achieves this evolution using the formula
\begin{equation}
\Omega(t)=\frac{\ddot{\chi}}{2\sqrt{\frac{\Delta^{2}}{4}-\dot{\chi}^{2}}%
}-\sqrt{\frac{\Delta^{2}}{4}-\dot{\chi}^{2}}\cot(2\chi), \label{Omegafromchi}%
\end{equation}
where $\Delta$ is the detuning of the pulse relative to the harmful
transition. Since the pulse is chosen to be resonant with the target
transition, we have $\Delta=\delta$, where $\delta$ is the detuning between
the target and harmful transitions. In Ref.~[\onlinecite{Economou_PRB15}], it
was shown that achieving an identity operation on the harmful transition
requires that the following conditions be satisfied: $\chi(0)=0$, $\dot{\chi
}(0)=0$, $\chi(\tau_{p})=\pi/4$, $\dot{\chi}(\tau_{p})=0$, $|\dot{\chi}(t)|\leq|\frac{\delta
}{2}|$, and $\psi_{\pm}%
(\tau_{p})=\frac{\delta\tau_{p}}{2}$ where $\psi_{\pm}(t)=\int_0^t dt'\sqrt{\Delta^2/4-\dot\chi^2(t')}\csc[2\chi(t')]\pm\tfrac{1}{2}\hbox{arcsin}(2\dot\chi(t)/\Delta)$. A choice of $\chi(t)$ satisfying these conditions was found to be
\begin{equation}
\chi(t)=A(t/\tau_{p})^{4}(1-t/\tau_{p})^{4}+\pi/4, \label{chi}%
\end{equation}
with $A=138.9$, and where the pulse duration is $\tau_{p}=5.87/|\delta|$. The
pulse shape that results from plugging Eq.~\eqref{chi} into
Eq.~\eqref{Omegafromchi} is shown in Fig. 1.

\begin{figure*}[t]
\centering
\includegraphics[width=1\columnwidth]{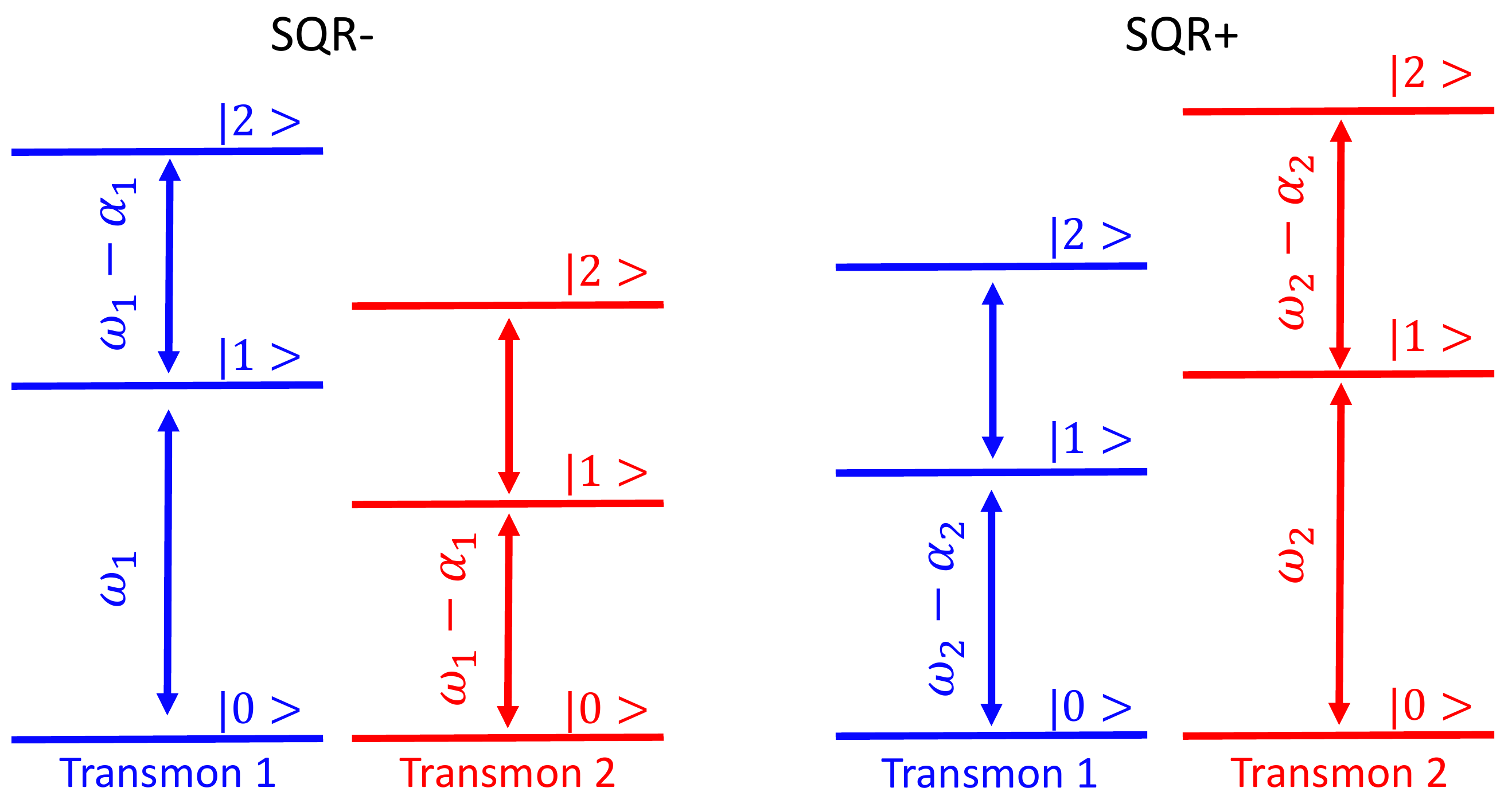}\caption{Diagram of bare
(non-interacting) energy levels for the two secondary qubit resonances (SQR)
at which the $\left|  0\right>  \Leftrightarrow\left|  1\right>  $ transition
of one qubit is resonant with the $\left|  1\right>  \Leftrightarrow\left|
2\right>  $ transition of the other. These resonances give rise to the bright
linear off-diagonal features evident in Fig.~\ref{Fig_FidelityQEn}(a).}
\label{Fig_sqr1}
\end{figure*}
\begin{figure*}[ptb]
\centering
\includegraphics[width=1\columnwidth]{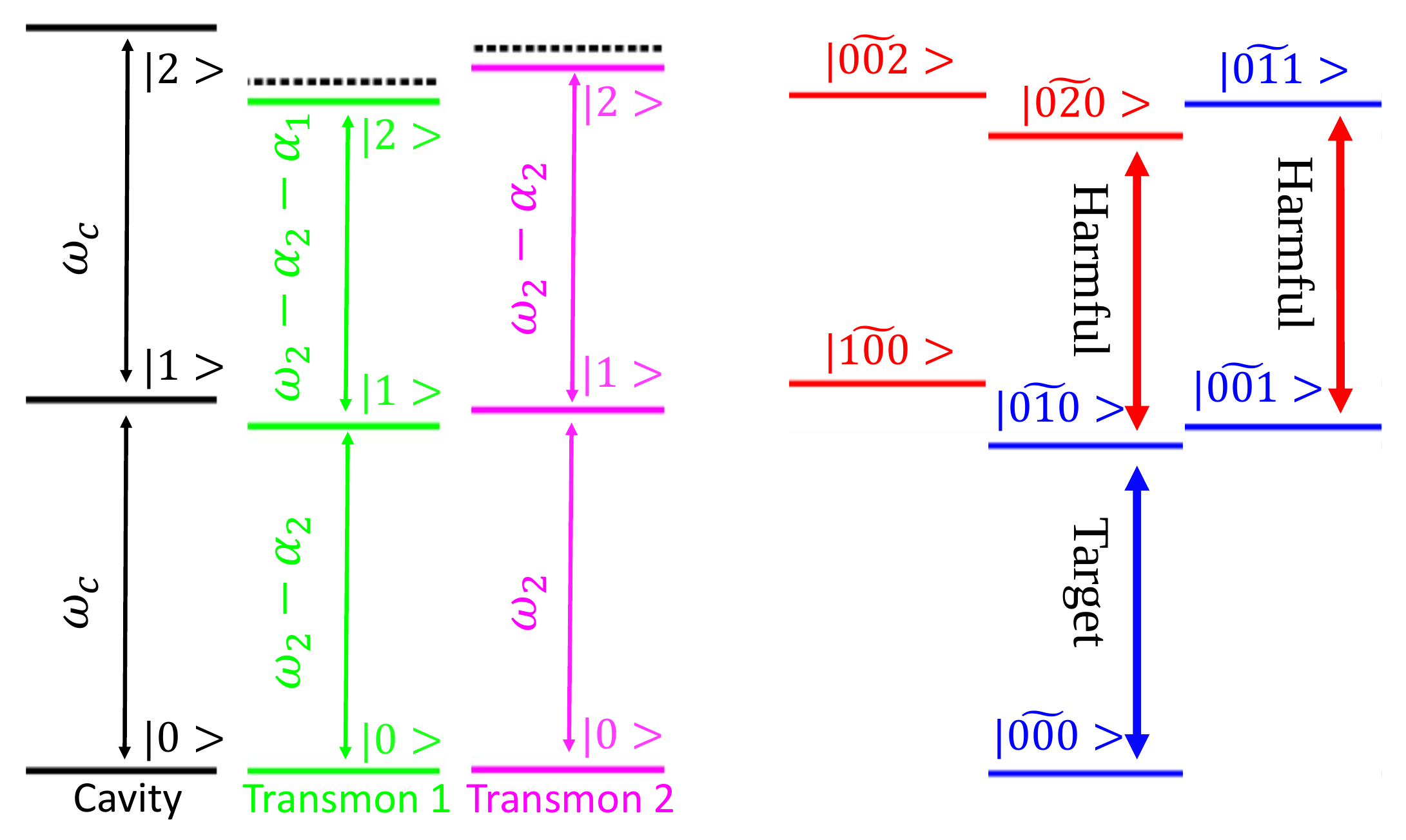}\caption{Dressed state
energy level diagram for the secondary qubit resonance at which the
$\left\vert 0\right\rangle \Leftrightarrow\left\vert 1\right\rangle $
transition of qubit 2 is resonant with the $\left\vert 1\right\rangle
\Leftrightarrow\left\vert 2\right\rangle $ transition of qubit 1, which is
driven. The degeneracy between the bare states $\left\vert 011\right\rangle $
and $\left\vert 020\right\rangle $ at this resonance leads to a large mixing
of these states when interactions are turned on, creating a large splitting
between the dressed states $\protect\widetilde{\left\vert 011\right\rangle }$
and $\protect\widetilde{\left\vert 020\right\rangle }$. This in turn leads to
a large detuning between the target and harmful transitions and hence a
reduced gate time, which is inversely proportional to the target-harmful
detuning under the SWIPHT protocol.}%
\label{Fig_sqr2}%
\end{figure*}

\section{Analysis of secondary resonances}\label{app:secondaryresonances}

In this appendix, we provide a more detailed analysis of the secondary
resonances, near which optimal gate performance can
be achieved. We elucidate the origin of the gate speed-up near the secondary
qubit resonances that is evident in Fig.~\ref{Fig_FidelityQEn}. At these
resonances, the $\left\vert 0\right\rangle \Leftrightarrow\left\vert
1\right\rangle $ transition of one qubit is resonant with the $\left\vert
1\right\rangle \Leftrightarrow\left\vert 2\right\rangle $ transition of the
other (see Fig.~\ref{Fig_sqr1}).

For concreteness, we focus on the secondary qubit resonance at which the
$\left|  0\right>  \Leftrightarrow\left|  1\right>  $ transition of qubit 2 is
resonant with the $\left|  1\right>  \Leftrightarrow\left|  2\right>  $
transition of qubit 1, which is driven. The degeneracy between the bare states
$\left|  011\right>  $ and $\left|  020\right>  $ at this resonance leads to a
large mixing of these states when interactions are turned on. This gives rise
to a large splitting between the dressed states $\widetilde{\left|
011\right>  }$ and $\widetilde{\left|  020\right>  }$, and in particular the
state $\widetilde{\left|  011\right>  }$ gets pushed to an energy that is
higher than what it would be further away from the resonance (see
Fig.~\ref{Fig_sqr2}). This means that the detuning between the target
transition, $\widetilde{\left|  000\right>  }\Leftrightarrow\widetilde{\left|
010\right>  }$, and the harmful transition, $\widetilde{\left|  001\right>
}\Leftrightarrow\widetilde{\left|  011\right>  }$, becomes larger. Since in
the SWIPHT protocol gate time is inversely proportional to this detuning, the
gate time is reduced near this secondary resonance. This is evident in Figs.~\ref{Fig_FidelityQEn}(b-d).

\section{Numerical simulations using parameters from experimental circuits}\label{app:experimentalparameters}

\begin{table*}[t]
\centering%
\begin{tabular}
[c]{|c|c|c|c|c|c|c|c|}\hline
\textbf{Reference} & \textbf{IBM} & \textbf{NIST} & \textbf{Yale} & Yale$^{-}$
& \textbf{Delft} & \textbf{ETH} & \textbf{LPS}\\\hline
$\mathbf{\omega}_{c}$(GHz$\times2\pi$) & $6.31$ & $5.7^{\ast}$ & $7.5$ & $7.5$
& $6.8478$ & $7.348$ & $6.6$\\\hline
$\mathbf{\omega}_{1}$(GHz$\times2\pi$) & $4.917$ & $4.72$ & $6.5^{\ast}$ &
$6.5^{\ast}$ & $\mathit{5.8899}$ & $\mathit{6.18}$ & $5.6$\\\hline
$\mathbf{\omega}_{2}$(GHz$\times2\pi$) & $5.415$ & $5.1$ & $6.18$ & $6.18$ &
$6.477$ & $7.0335$ & $\mathit{6.3}$\\\hline
$\mathbf{\alpha}_{1},\mathbf{\alpha}_{2}$(GHz$\times2\pi$) & $330,330^{\ast}$
& $284$ & $209$ & $209$ & $\mathit{350}$ & $300^{\ast}$ & $211$\\\hline
$\mathbf{g}_{1}\mathbf{,g}_{2}$(MHz$\times2\pi$) & $250^{\ast}$ & $125$ &
$250^{\ast}$ & $250^{\ast}$ & $250^{\ast}$ & $250^{\ast}$ & $250^{\ast}%
$\\\hline
$\mathbf{T}_{1}$($\mu$s) & $30$ & $\mathit{20}$ & $60$ & $15$ & $25$ & $1.33$
& $24$\\\hline
$\mathbf{T}_{2}$($\mu$s) & $13.8$ & $\mathit{20}$ & $8.4$ & $8.4$ & $39$ &
$0.967$ & $41$\\\hline
$\mathcal{F}^{+}$ & 0.9936 & 0.9760 & 0.9951 & 0.9937 & 0.9942 & 0.8503 &
0.9900\\\hline
$\tau_{g}^{+}$\textbf{(ns)} & 57.0389 & 167.5833 & 41.8092 & 41.5766 &
35.9051 & 73.1032 & 73.6202\\\hline
$\Delta^{+}$\textbf{(MHz}$\times2\pi$\textbf{)} & 16.3790 & 5.5748 & 22.3453 &
22.4986 & 26.0197 & 12.7797 & 12.6900\\\hline
$\mathcal{F}^{ideal}$ & 0.9996 & 0.9998 & 0.9981 & 0.9979 & 0.9983 & 0.9994 &
0.9987\\\hline
$\tau_{g}^{ideal}$\textbf{(ns)} & 57.0389 & 172.5122 & 41.8092 & 41.5766 &
35.8913 & 83.0718 & 73.6202\\\hline
$\Delta^{ideal}$\textbf{(MHz}$\times2\pi$\textbf{)} & 16.3790 & -5.4155 &
22.3453 & 22.4986 & 25.7146 & 11.2462 & -12.6900\\\hline
$T_{1}^{\mathcal{F}=0.999}$($\mu s$) & 92.5 & 175.08 & N/A & N/A & N/A &
164.5 & N/A\\\hline
\end{tabular}
\caption{Experimental parameters and our simulation results. Numbers with
asterisks indicate values that have been modified to yield improved results.
Numbers in italics indicate values that were taken from other works since they
were not provided in the paper. \{$\mathcal{F}^{+}$, $\tau_{g}^{+}$,
$\Delta^{+}$\} are the results for fidelity, gate time, and pulse detuning,
respectively, obtained by adjusting one of the qubit frequencies for improved
performance. \{$\mathcal{F}^{ideal}$, $\tau_{g}^{ideal}$, $\Delta^{ideal}$\}
are the results for the fidelity in the absence of relaxation and dephasing
for the improved parameters. The column labeled Yale$^{-}$ accounts for a
reduction in relaxation time $T_{1}$ as a consequence of the enhancement in
cavity-qubit coupling $g_{1,2}$.}%
\label{Table_ExpDat2}%
\end{table*}
\begin{table*}[t]
{\tiny \centering%
\begin{tabular}
[c]{|c|c|c|c|c|c|c|c|c|c|c|c|c|}\hline
\textbf{Reference} & \textbf{IBM} & \textbf{IBM}$^{D2Q}$ & \textbf{NIST} &
\textbf{NIST}$^{D2Q}$ & \textbf{Yale} & \textbf{Yale}$^{D2Q}$ & \textbf{Delft}
& \textbf{Delft}$^{D2Q}$ & \textbf{ETH} & \textbf{ETH}$^{D2Q}$ & \textbf{LPS}
& \textbf{LPS}$^{D2Q}$\\\hline
$\mathbf{\omega}_{c}$(GHz$\times2\pi$) & $6.494$ & $6.494$ & $6.3(5.6^{\ast})$
& $6.3(5.6^{\ast})$ & $7.5$ & $7.5$ & $6.8506$ & $6.8506$ & $7.347$ & $7.347$
& $6.6$ & $6.6$\\\hline
$\mathbf{\omega}_{1}$(GHz$\times2\pi$) & $4.917(4.72^{\ast})$ &
$4.917(4.72^{\ast})$ & $4.72$ & $4.72$ & $4.87(6.5^{\ast})$ & $6.5(6.5^{\ast
})$ & $\mathit{5.8899}$ & $\mathit{5.8899}$ & $\mathit{6.18}$ & $\mathit{6.18}%
$ & $5.6$ & $5.6$\\\hline
$\mathbf{\omega}_{2}$(GHz$\times2\pi$) & $5.415$ & $5.415$ & $5.1$ & $5.1$ &
$6.18$ & $6.18$ & $6.477$ & $6.477$ & $7.0335$ & $7.0335$ & $\mathit{6.3}$ &
$\mathit{6.3}$\\\hline
$\mathbf{\alpha}$(GHz$\times2\pi$) & $330$ & $330$ & $284$ & $284$ & $212$ &
$212$ & $\mathit{350}$ & $\mathit{350}$ & $90(300^{\ast})$ & $90(300^{\ast})$
& $211$ & $211$\\\hline
$\mathbf{g}_{1}\mathbf{,g}_{2}$(MHz$\times2\pi$) & $250^{\ast}$ & $250^{\ast}$
& $125$ & $125$ & $250^{\ast}$ & $250^{\ast}$ & $250^{\ast}$ & $250^{\ast}$ &
$250^{\ast}$ & $250^{\ast}$ & $70(250^{\ast})$ & $250^{\ast}$\\\hline
$\mathbf{T}_{1}$($\mu$s) & $30$ & $30$ & $\mathit{20}$ & $\mathit{20}$ & $60 $
& $60$ & $25$ & $25$ & $1.33$ & $1.33$ & $24$ & $24$\\\hline
$\mathbf{T}_{2}$($\mu$s) & $13.8$ & $13.8$ & $\mathit{20}$ & $\mathit{20}$ &
$8.4$ & $8.4$ & $39$ & $39$ & $0.967$ & $0.967$ & $41$ & $41$\\\hline
$\mathcal{F}$ & 0.9925 & 0.9910 & 0.9293 & 0.9334 & 0.8957 & 0.8870 & 0.9942 &
0.9981 & 0.6756 & 0.6440 & 0.6430 & \\\hline
$\mathbf{\tau}_{g}$(ns) & 69.5868 & 61.7387 & 511.2219 & 473.2454 & 1512.3 &
1498.8 & 35.8913 & 35.8913 & 200.6993 & 200.6993 & 4328.7 & \\\hline
$\Delta$\textbf{(MHz}$\times2\pi$\textbf{)} & 13.4255 & 15.1322 & 1.8275 &
1.9741 & 0.6177 & 0.6233 & 25.7146 & 26.0297 & 4.6549 & 4.6549 & 0.2158 &
\\\hline
$\mathcal{F}^{\ast}$ & 0.9821 & 0.9828 & 0.9760 & 0.9741 & 0.9949 & 0.9555 &
0.9942 & 0.9981 & 0.8492 & 0.8586 & 0.9904 & 0.9904\\\hline
$\tau_{g}^{\ast}$\textbf{(ns)} & 173.2139 & 168.9891 & 167.5833 & 186.1 &
41.1427 & 42.8220 & 35.8913 & 35.8913 & 72.7194 & 73.1099 & 73.6202 &
69.9620\\\hline
$\Delta^{\ast}$\textbf{(MHz}$\times2\pi$\textbf{)} & 5.3936 & 5.5284 &
5.5748 & 5.0199 & 22.7073 & 21.8168 & 25.7146 & 26.0297 & 12.8472 & 12.7786 &
12.6900 & 13.3535\\\hline
$\mathcal{F}^{\ast ideal}$ & 0.9999 &  & 0.9998 &  & 0.9987 & 0.9972 &
0.9983 &  & 0.9993 &  & 0.9988 & 0.9977\\\hline
$\tau_{g}^{\ast ideal}$\textbf{(ns)} & 173.2139 &  & 172.5122 &  & 41.1427 &
40.8755 & 36.3311 &  & 83.4134 &  & 72.7289 & 69.9620\\\hline
$\Delta^{\ast ideal}$\textbf{(MHz}$\times2\pi$\textbf{)} & -5.3936 &  &
5.4155 &  & 22.7073 & 22.8557 & 25.7146 &  & 11.2001 &  & 12.8455 &
13.3535\\\hline
$T_{1}^{\mathcal{F}=0.999}$($\mu s$) & 166.25 &  & 175.08 &  & N/A &  & N/A &
& 175 &  & N/A & \\\hline
\end{tabular}
}\caption{SWIPHT \textsc{cnot} gate performance in the case where both qubits
are driven (D2Q) at the same time. Numbers with asterisks indicate values that
have been modified to yield improved results. Numbers in italics indicate
values that were taken from other works since they were not provided in the
paper.}%
\label{Table_ExpDatomega2}%
\end{table*}
We examine the performance for several sets of parameters taken from
experimental works and indicate ways to further improve results through
minimal parameter adjustments. We have simulated the SWIPHT \textsc{cnot} gate
performance using parameters extracted from experimental works, including
those of IBM,\cite{McKay_PRApplied16,Sheldon_PRA16,Corcoles_NC15} NIST,
\cite{Private} Yale,\cite{Liu_PRX16} Delft,\cite{Bultink_arxiv16} ETH,
\cite{Berger_NC15} LPS, \cite{Private} as shown in the following tables. We
have optimized the fidelity over the pulse frequency for each row of data. Asterisks indicate parameters that have been adjusted relative to
what was used in the corresponding paper in order to improve performance. We
have increased the coupling in cases where the cavity was too far detuned from
the qubits to yield feasible gate times within the SWIPHT scheme. In general,
SWIPHT works when the degeneracy between the target and harmful transitions is
strongly broken, which requires either strong qubit-cavity couplings, reduced
qubit-cavity detunings, or tuning qubit parameters to lie near secondary
resonances (see Appendix~\ref{app:secondaryresonances}). Fidelities outside the operational regime of
SWIPHT are typically below 90\%. Couplings up to $250$ MHz are experimentally
reasonable since there exist experimental filtering techniques that can enable
one to increase the coupling strength without sacrificing $T_{1}$ times
through Purcell effects.\cite{Reed_APL10,Whittaker_PRB14, Bronn_IEEE15} In
the column labeled Yale$^{-}$ in Table \ref{Table_ExpDat2}, we show the
performance without such filtering, where the relaxation time is reduced by a
factor of 4 as a consequence of the factor of 2 enhancement in qubit-cavity
coupling. We see that the performance is not significantly affected provided
the original relaxation time is well above 10 $\mu$s. As described in the main
text and in Appendix~\ref{app:secondaryresonances}, we have demonstrated a way to
improve the gate quality \{$\mathcal{F}^{+}$, $\tau_{g}^{+}$, $\Delta^{+}$\}
by tuning one qubit frequency so that the system lies near a secondary
resonance. \{$\mathcal{F}^{ideal}$, $\tau_{g}^{ideal}$, $\Delta^{ideal}$\} are
the results for the fidelity (obtained from quantum state tomography) without
noise for the improved parameters. In the last row, $T_{1,2}^{F=0.999}$
indicates a threshold of decoherence in order to reach a fidelity of 99.9\%
for a specific set of parameters with corresponding $\Delta^{+}$. Here we have
assumed $T_{1}^{0.999}=T_{2}^{0.999}/2$. This threshold $T_{1}$ value provides
an idea of the noise level needed for a specific transmon system to achieve
$0.999$ fidelity for a \textsc{cnot} gate based on our scheme. Table
\ref{Table_ExpDat2} shows that it is possible to obtain fidelities in excess
of 0.99 while keeping pulse times below 100 ns in most cases even with
realistic noise included. The ideal fidelity values further show that most of
the residual gate error is caused by decoherence and relaxation. Table
\ref{Table_ExpDatomega2} gives similar results for additional parameter sets.
The table further shows that the results are essentially the same when the
driving is allowed to act on both qubits.

\begin{figure}[ptb]
\centering
\includegraphics[width=0.8\columnwidth]{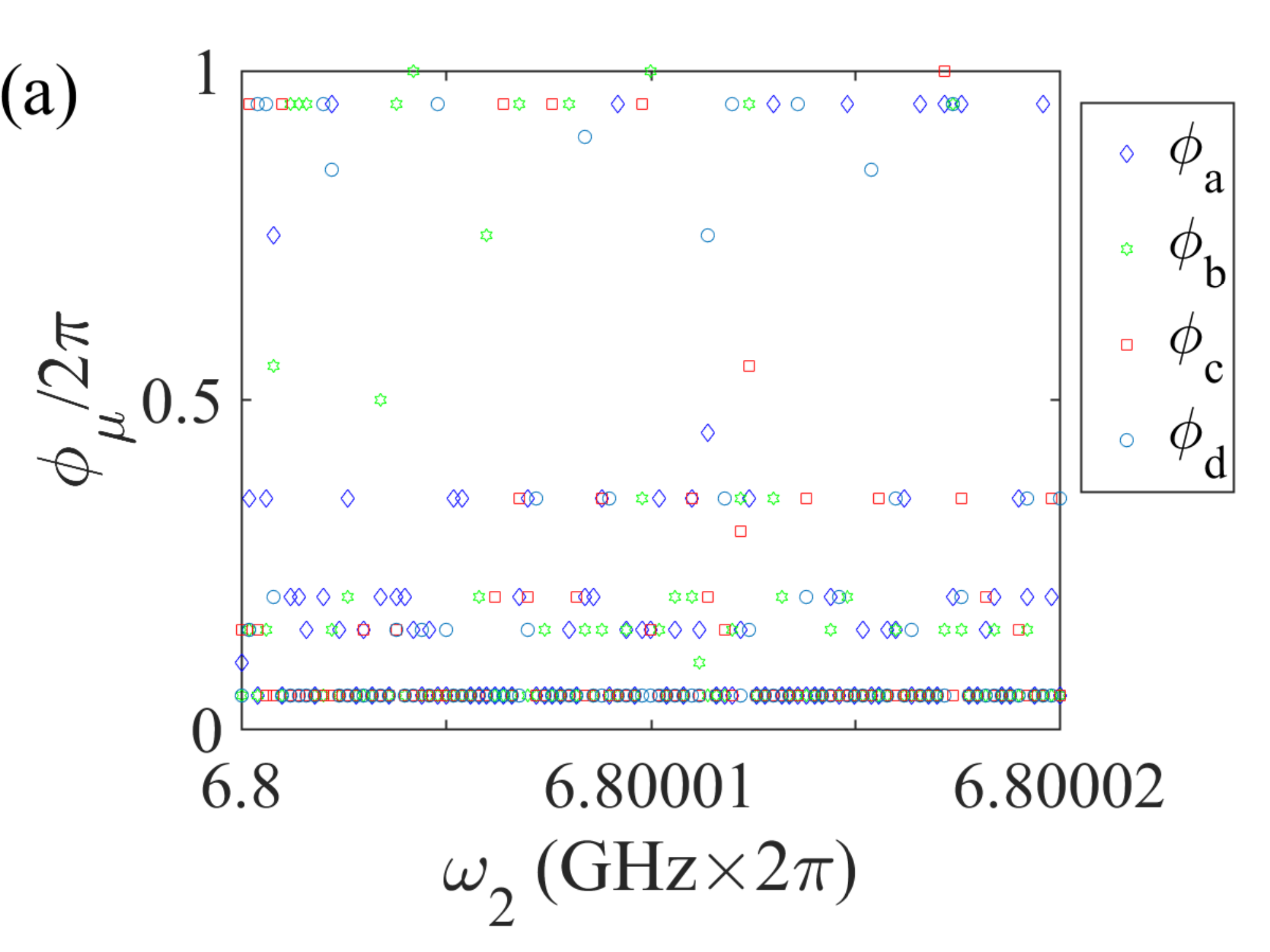}
\includegraphics[width=0.8\columnwidth]{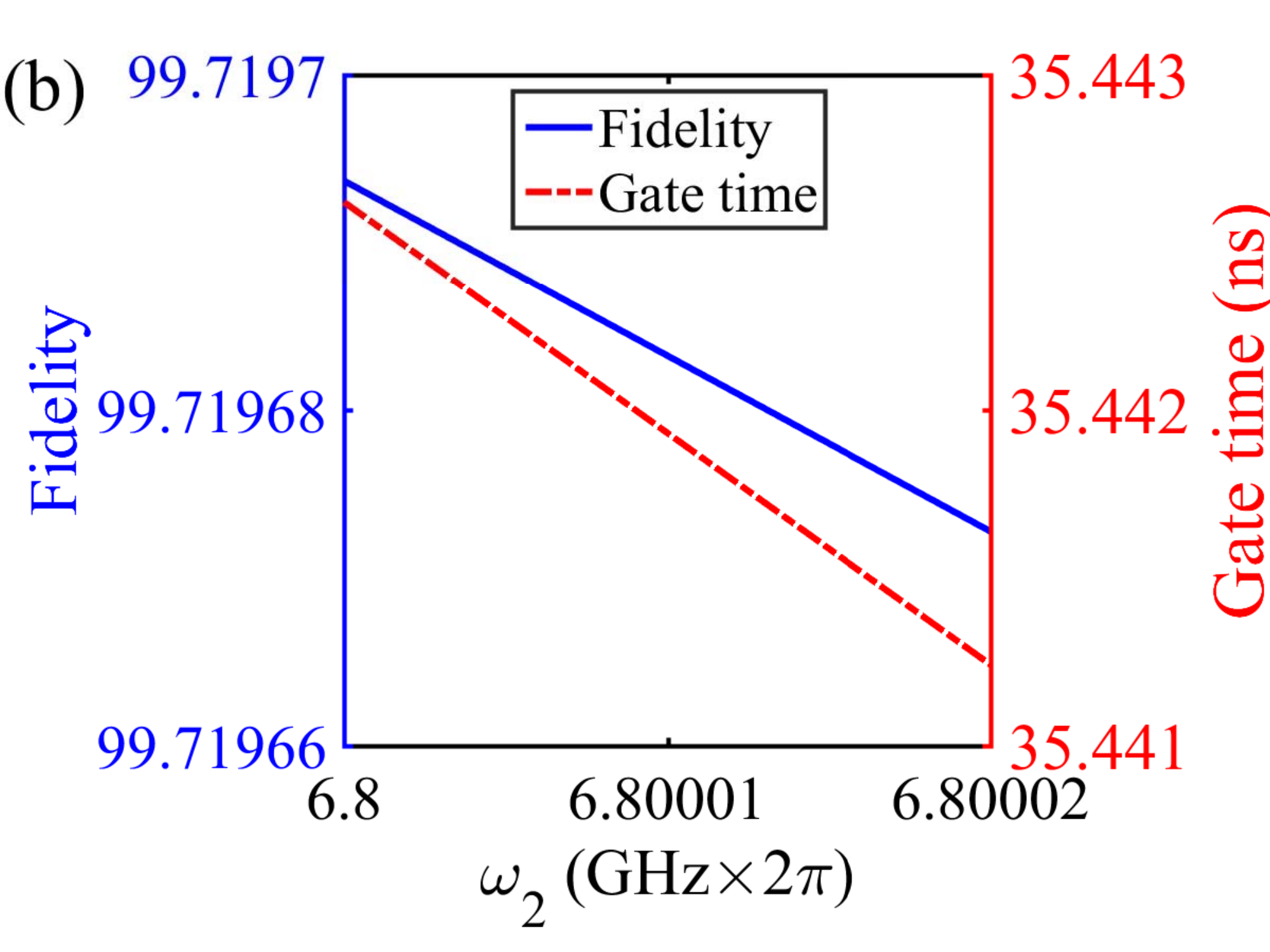}\caption{(a)
Local phases of the generalized \textsc{cnot} as a function of qubit
frequency. (b) Insensitivity of the SWIPHT gate fidelity with respect to local
phases. The fidelity as a function of qubit frequency is shown for fixed
values of local phases. The system parameters are as in Fig.~\ref{fig:pulse} and $T_{1}=T_{2}=20\mu s$.}%
\label{Fig_fid_phase}%
\end{figure}

\section{Sensitivity to local phases of the generalized CNOT}\label{app:localphases}

We consider how the phases entering into the definition of our generalized
CNOT gate, Eq. (\ref{cnot}), depend on system parameters.
These phases represent the trivial, local part of the entangling gate and can
be corrected with local single-qubit gates. Due to the finite linewidths of
the transmon excited states, there exists experimental uncertainty in the
values of the transmon frequencies (on the order of 10 kHz), and this can in
turn create uncertainty in the values of the local phases. In Fig.
\ref{Fig_fid_phase} we show that although the local phases are sensitive to
qubit frequencies (Fig. \ref{Fig_fid_phase} (a)), the SWIPHT CNOT gate
fidelity remains essentially constant as qubit frequencies are varied over a
range of $20$ kHz even when the local phases are held fixed, demonstrating
that the gate performance is not sensitive to these phases or to typical
levels of uncertainty in qubit frequencies (Fig. \ref{Fig_fid_phase} (b)).

\end{document}